\documentclass[authoryear, 12]{elsarticle}
\usepackage{multirow}
\usepackage{amstext}
\usepackage{amssymb}
\usepackage{graphicx}

\makeatletter

\usepackage{hyperref}
\usepackage{lineno}

\makeatother

\begin{document}

\begin{frontmatter}{}

\title{Realistic uncertainties on Hapke model parameters from photometric
measurement}

\author{Schmidt Fr{\'e}d{\'e}ric$^{1,2}$, Fernando Jennifer$^{1,2}$}

\address{$^{1}$ Univ. Paris-Sud, Laboratoire GEOPS, UMR8148, Orsay, F-91405;
(frederic.schmidt@u-psud.fr) $^{2}$ CNRS, Orsay, F-91405, France}
\begin{abstract}
The single particle phase function describes the manner in which an
average element of a granular material diffuses the light in the angular
space usually with two parameters: the asymmetry parameter $b$ describing
the width of the scattering lobe and the backscattering fraction $c$
describing the main direction of the scattering lobe. Hapke proposed
a convenient and widely used analytical model to describe the spectro-photometry
of granular materials. Using a compilation of the published data,
Hapke (2012, Icarus, 221, 1079-1083) recently studied the relationship
of $b$ and $c$ for natural examples and proposed the hockey stick
relation (excluding $b>0.5$ and $c>0.5$). For the moment, there
is no theoretical explanation for this relationship. One goal of this
article is to study a possible bias due to the retrieval method.

We expand here an innovative Bayesian inversion method in order to
study into detail the uncertainties of retrieved parameters. On Emission
Phase Function (EPF) data, we demonstrate that the uncertainties of
the retrieved parameters follow the same hockey stick relation, suggesting
that this relation is due to the fact that $b$ and $c$ are coupled
parameters in the Hapke model instead of a natural phenomena. Nevertheless,
the data used in the Hapke (2012) compilation generally are full Bidirectional
Reflectance Diffusion Function (BRDF) that are shown not to be subject
to this artifact.

Moreover, the Bayesian method is a good tool to test if the sampling
geometry is sufficient to constrain the parameters (single scattering
albedo, surface roughness, $b$, $c$, opposition effect). We performed
sensitivity tests by mimicking various surface scattering properties
and various single image-like/disk resolved image, EPF-like and BRDF-like
geometric sampling conditions. The second goal of this article is
to estimate the favorable geometric conditions for an accurate estimation
of photometric parameters in order to provide new constraints for
future observation campaigns and instrumentations.\end{abstract}

\end{frontmatter}{}


\section{Introduction}

Photometry is the study of the surface by the angular response of
the reflected light by a medium described by the Bi-directional Reflectance
Distribution Function (BRDF) \citep{Hapke_Book1993}.

Hapke proposed a set of approximated analytical equations to estimate
conveniently the BRDF of a granular medium \citep[e.g.][]{Hapke_BRDF1theroy_JGR1981,Hapke_BRDF2Experiments_JGR1981,Hapke_BRDF3roughness_Icarus1984,Hapke_BRDF4ExtinctionOpposition_Icarus1986,Hapke_BRDF5scatterringCBOE_Icarus2002,Hapke_BRDF6_Porosity_Icarus2008}.
This formulation has been controversial for two decades \citep[e.g.][]{Mishchenko1994,Hapke_PlanetaryBackscatering_JQSRT1996,Shepard_testHapkephotometric_JoGRP2007,Shkuratov_criticalassessmentHapke_JoQSaRT2012,Hapke_CommentAcritical_JoQSaRT2013}
but due to its relative simplicity and fast computation, many authors
have been using it to analyze laboratory data \citep[e.g.][]{Cord_PlanetBDRF_Icarus2003,Souchon_experimentalstudyof_I2011,Beck_Photometryofmeteorites_I2012,Pommerol_PhotometricpropertiesMars_JGRP2013,Johnson_Spectrogoniometryandmodeling_I2013},
telescopic observation \citep[e.g.][]{hapke_oppostition_Icarus1998},
in situ data \citep[e.g.][]{Johnson_Preliminaryresultsphotometric_JGR1999,Johnson_Spirit_JGR2006,Johnson_Opportunity_JGR2006},
remote sensing data \citep[e.g.][]{Jehl_GusevPhotometry_Icarus2008,Yokota_Lunarphotometricproperties_I2011,Fernando_SurfacereflectanceMars_JoGRP2013,Vincendon_Marssurfacephase_PaSS2013,Sato_ResolvedHapkeparameter_JoGRP2014}.

The scope of this article is to discuss the properties of the Hapke
model in term of data analysis, but not to discuss particular aspects
of the realism of the photometric Hapke model. Some authors addressed
the difficulties to fit the model to actual data, indicating that
the problem is ill posed, due to parameter coupling \citep{Mustard_Photometry_JGR1989,1989aste.conf..557H,2006LPI....37.1340B,Souchon_experimentalstudyof_I2011}.
The most usual ways to fit are a minimization of the $\chi^{2}$,
stepwise in a grid \citep{Shepard_testHapkephotometric_JoGRP2007},
using the Levenberg-Marquardt method \citep{Sato_ResolvedHapkeparameter_JoGRP2014,Johnson_Spectrogoniometryandmodeling_I2013},
a simplex algorithm \citep{Gunderson_Firstmeasurementswith_PaSS2006},
and a genetic algorithm \citep{Cord_PlanetBDRF_Icarus2003} and Particle
Swarm Optimization \citep{Beck_Photometryofmeteorites_I2012,Pommerol_PhotometricpropertiesMars_JGRP2013}.
These strategies are relevant in the case of an unique solution (only
one minimum in the $\chi^{2}$) and close to gaussian shape around
the solution. Due to the non-linearities of the Hapke equations, those
two mathematical properties may not be fulfilled.

We propose here a new kind of technique, based on the Bayesian formulation
to estimate the model parameters in agreement with the data \citep{Tarantola_QuestInformation_JGeophys1982}.
The general framework of Bayesian theory enabled us to: (i) take into
account precisely the uncertainties of measured quantities, (ii) define
precisely the range of possibility of all model parameters, (iii)
estimate the range of solution in a general case (that may not have
a gaussian shape). From theoretical point of view, each information
is described as a probability density function (PDF): the measured
quantities, the a priori parameters and the posterior parameters.
Within this framework, the solution is expressed as a ``final state
of information'' which always exists, solving the apparent ill-posed
problem.

We already used this strategy to analyze the spectro-photometric data
from the Compact Reconnaissance Imaging Spectrometer for Mars (CRISM)
instrument \citep{Murchie_CompactReconnaissanceImaging_JGR2007},
after the MARS-ReCO atmospherical correction \citep{Ceamanos_SurfacereflectanceMars_JoGRP2013},
by estimating the PDF of all Hapke parameters of the Martian surface
\citep{Fernando_SurfacereflectanceMars_JoGRP2013}. We also applied
it on full CRISM images on the MER landing sites \citep{Fernando_Characterizationandmapping_I2015}. 

In this article, we propose to perform extensive sensitivity tests
on synthetical dataset, using the Bayesian inversion, in order to
study the behavior of Hapke model to fit the data. This study aims
at: 
\begin{enumerate}
\item estimating precise uncertainties level on the model parameters on
different typical observation types (one image-like observation/disk
resolved image, CRISM-like Emission Phase Function (EPF), very favorable
EPF, Bidirectional Reflectance Distribution Function (BRDF) measurements)
\item determining the uncertainties dependance on BRDF sampling
\item evaluating the possibility to have more than one ``solution'', i.e.
multiple minima
\item chasing any effect which could bias the estimation of the parameters
$b$ and $c$ leading to a fake hockey stick relationship
\end{enumerate}
This work should help to interpret previous analyses but also to design
future instrumental and observational campaigns.

\section{Method}

\subsection{Direct model: the Hapke's photometric model \label{sub:Direct-model}}

Standard 1993 Hapke modeling \citep{Hapke_Book1993} is widely used
in the planetary science community due to the simplicity of its expression,
fast computation, and the purported physical meaning of the model
parameters allowing the characterization of planetary surface materials
(e.g., grain size, morphology, internal structure and surface roughness). 

More recent developments are available. First, the version from \citealp{Hapke_BRDF5scatterringCBOE_Icarus2002}
includes: (i) a more accurate analytic approximation for isotropic
scatterers, (ii) a better estimation of the bidirectional reflectance
when the scatterers are anisotropic, and (iii) the incorporation of
coherent backscattering. Second, the version from \citealp{Hapke_BRDF6_Porosity_Icarus2008}
treats the porosity. 

Following previous studies \citep{Johnson_Spirit_JGR2006,Johnson_Opportunity_JGR2006,Jehl_GusevPhotometry_Icarus2008,Beck_Photometryofmeteorites_I2012,Fernando_SurfacereflectanceMars_JoGRP2013},
we decided to use the following standard expression of Hapke's 1993
in order to be coherent with older analysis. In addition, more recent
developments are not fully validated with experimental data. We remind
here the main expression:

\begin{equation}
r\left(\theta_{0},\theta,g\right)=\frac{\omega}{4\pi}\,\frac{\mu_{0e}}{\left(\mu_{0e}+\mu_{e}\right)}\,\left\{ \left[1+B(g)\right]P(g)+H(\mu_{0e})H(\mu_{e})-1\right\} S(\theta_{0},\theta,g)\label{eq:Hapke}
\end{equation}

Using the following quantities:
\begin{itemize}
\item $\theta_{0}$, $\theta$, and $g$: incidence, emergence and phase
angles. The whole geometry quantities are noted $\Omega=(\theta_{0},\theta,g)$.
Terms $\mu_{0e}$ and $\mu_{e}$ are the cosine of the equivalent
incidence and emergence angles, in the case of a rough surface, as
defined in \citep{Hapke_Book1993}. We note $\phi$ as the azimuth
angle.
\item $\omega$ ($0\le\omega\le1$): single scattering albedo. It represents
the fraction of scattered to incident radiation by a single particle
(also noted w).
\item $P(g)$: particle scattering phase function. It characterizes the
angular distribution of energy for an average particle. We used the
empirical 2-term Henyey-Greenstein function \citep{Henyey_DiffusionGalaxy_ApJ1941}
(hereafter referred to as HG2) for studying planetary surfaces \citep{Cord_PlanetBDRF_Icarus2003,Jehl_GusevPhotometry_Icarus2008,Johnson_Spirit_JGR2006,Johnson_Opportunity_JGR2006,Souchon_experimentalstudyof_I2011,Beck_Photometryofmeteorites_I2012,Pommerol_PhotometricpropertiesMars_JGRP2013}:
\begin{equation}
P(g)=\left(1-c\right)\,\frac{1-b^{2}}{\left(1+2b\cos\left(g\right)+b^{2}\right)^{3/2}}+c\,\frac{1-b^{2}}{\left(1-2b\cos\left(g\right)+b^{2}\right)^{3/2}}\label{eq:HG2}
\end{equation}
The HG2 function depends on two parameters: the asymmetry parameter
$b$ ($0\le b\le1$) which characterizes the anisotropy of the scattering
lobe (from $b=0$, which corresponds to the isotropic case, to $b=1$,
which corresponds to a particle which diffuses light in a single direction)
and the backscattering fraction $c$ ($0\le c\le1$) which characterizes
the main direction of the diffusion ($c<0.5$ corresponding to forward
scattering and $c>0.5$ corresponding to backward scattering).
\item $H(x)$: multiple scattering function. An analytical function for
isotropic scatterers has been proposed \citep{Hapke_Book1993} with
a relative error to the exact values \citep{Chandrasekhar_Book1960}
lower than 1\%, leading to a relative error on the BRDF lower than
2\% \citep{Cheng_RadiativeTransfert_JGR2000}. Defining $y=(1-\omega)^{1/2}$,
the multiple scattering function is:
\begin{equation}
H(x)=\left\{ 1-\left[1-y\right]x\left[\left(\frac{1-y}{1+y}\right)+\left(1-\frac{1}{2}\left(\frac{1-y}{1+y}\right)-x\left(\frac{1-y}{1+y}\right)\right)\ln\left(\frac{1+x}{x}\right)\right]\right\} ^{-1}\label{eq:H1993}
\end{equation}
A new expression dedicated to anisotropic scattering has been proposed
\citep{Hapke_BRDF5scatterringCBOE_Icarus2002}. Nevertheless, \citet{Pommerol_PhotometricpropertiesMars_JGRP2013}
have noticed that the use of the recent $H$ expression leads to no
significant changes over the previous expression.
\item $B(g)$: opposition effect function. It describes the sharp increase
of brightness around the zero phase angle often observed in the case
of particulate media. Only the Shadow Hiding Opposition Effect (SHOE)
is taken into account as follows \citep{Hapke_Book1993}:
\begin{equation}
B(g)=\frac{B_{0}}{1+\frac{1}{h}\tan\left(\frac{g}{2}\right)}
\end{equation}
The function depends on the parameters $h$ and $B_{0}$ (ranging
from 0 to 1) which are the angular width and the amplitude of the
opposition effect respectively. The Coherent Backscattering Opposition
Effect (CBOE) is ignored in our case since the minimum phase angle
is large ($g>10^\circ$).
\item $S(\theta_{0},\theta,g)$: macroscopic roughness factor. It describes
the surface topography as a set of facets with a Gaussian distribution
of tangent of the slopes, with mean slope angle noted $\bar{\theta}$
also called surface macroscopic roughness \citep{Hapke_Book1993}.
This model of the surface roughness effect includes the partially
shadowed area depending on the geometry and the bias on the effective
incidence and emergence angles. The expression of $S$ is given in
\citet{Hapke_Book1993}.
\end{itemize}

\subsection{Bayesian inversion\label{sub:Bayesian-inversion}}

Because the Hapke model is a non linear model and may not have a unique
solution, one may use the Bayesian inversion framework based on the
concept of the state of information which is characterized by a Probability
Density Function (PDF) \citep{Tarantola_QuestInformation_JGeophys1982}.
This approach has already been proposed for the Hapke model \citep{Fernando_SurfacereflectanceMars_JoGRP2013}.
To infer the solution, this approach takes into account the initial
state of information (\emph{a priori} knowledge) on the parameters
and the observations and applies the Bayes' theorem to estimate the
final state of information, called \emph{a posteriori}. The key points
of the concept and framework assumptions are presented in the following:
\begin{itemize}
\item Direct model and related quantities: 

\begin{itemize}
\item The direct model $F$ estimates the simulated data $d$, from model
parameters $m$: 
\begin{equation}
d=F(m)
\end{equation}

\item The parameters $m$ include $\omega$, $b$, $c$, $\bar{\theta}$,
$B_{0}$ and $h$. The parameters $m$ are in the parameter vector
space $M=[0,1]^{5}[0,45]$, since all parameters are between 0 and
1, except $\bar{\theta}$ between 0 and 45$^\circ$ .
\item The simulated data $d$ is the collection $r_{i}=r\left(\Omega_{i}\right)$
for all $i$th peculiar geometry noted as $\Omega_{i}=(\theta_{0},\theta,g)_{i}$.
The total number of geometries is noted $N_{g}$. The simulated data
$d$ are in the observation vector space $D=\mathbb{R}^{Nd+}$, since
the BRDF is a positive quantity.
\item The model $F(m)$ is the Hapke equation \ref{eq:Hapke}. We consider
that the geometries have very low uncertainties, which is the case
for most data in planetary science cases. Thus, the parameters $\Omega_{i}$
are not estimated by the inversion.
\end{itemize}
\item Observation and other \emph{a priori} information: 

\begin{itemize}
\item The actual observation is considered as prior information on data
$\rho_{D}(d)$ in the observation space $D$. It is assumed to be
a $N_{g}$-dimension gaussian PDF $\mathcal{G}(o,C)$, with mean $o$
and covariance matrix $C$. The values $o_{i}$ are the observation
for each geometry. The covariance matrix $C$ is assumed here to be
diagonal since each measurements at a given geometry is independent
of the other geometric measurements. The diagonal elements $C_{ii}$
are $\sigma_{1}^{2},\ldots,\sigma_{N_{g}}^{2}$, with $\sigma_{i}$
being the standard deviation.
\item The prior information on model parameters $\rho_{M}(m)$ in the parameter
space $M$ is independent to the data and corresponds to the state
of null information if no information is available on the parameters.
We consider an uniform PDF in their definition space $M$. Outside
$M$, the PDF is null, avoiding unphysical solutions to appear.
\item The state of null information $\mu_{D}(d)$, representing the case
when no information is available, is trivial in our case and represents
the uniform PDF in the parameters space $M$.
\end{itemize}
\item Solution of inverse problems and \emph{a posteriori} information: 

\begin{itemize}
\item The posterior PDF in the model space $\sigma_{M}(m)$ as defined by
\citet{Tarantola_QuestInformation_JGeophys1982} is: 
\begin{equation}
\sigma_{M}(m)=k\,\rho_{M}(m)\,L(m)
\end{equation}
where $k$ is a constant and $L(m)$ is the likelihood function
\begin{equation}
L(m)=\int_{D}\mathrm{d}d\,\frac{\rho_{D}(d)\,\theta(d\mid m)}{\mu_{D}(d)}
\end{equation}
where $\theta(d\mid m)$ is the theoretical relationship of the PDF
for $d$ given $m$. We do not consider errors on the model itself,
so $\theta(d\mid m)=\delta(F(m))$.
\item The solution of the general inverse problem is given by the PDF $\sigma_{M}(m)$.
The best way to represent $\sigma_{M}(m)$ is to plot the marginal
PDF $\sigma_{M}(m_{j})$ for one parameter $j$ (see for instance
fig. \ref{fig:Results_EPFexample_histSingle}), or the bivariate marginal
PDF $\sigma_{M}(m_{j},m_{j'})$ (see for instance fig. \ref{fig:Results_EPFexample_histDouble}).
\item The PDF can be described by statistic indicators such as mean values
(expectation), standard deviations (covariance matrix), higher order
statistics, etc. 
\end{itemize}
\end{itemize}

\subsection{Monte Carlo Markov Chain (MCMC) to sample the inverse problem\label{sub:MCMC}}

Since the Hapke model is non-linear, it is not possible to describe
the posterior PDF $\sigma_{M}(m)$ analytically. The solution provided
by \citet{Fernando_SurfacereflectanceMars_JoGRP2013} was to sample
the final solution using a Monte Carlo approach using the Metropolis
rule to built a Monte Carlo Markov chain (MCMC) \citep{Mosegaard_MonteCarloInversion_JGR1995}.
It consists in random testing of a candidate simulation over the parameter
space and keeping only some solution in order to follow the \emph{a
priori} information. After a sufficient number of steps, the chain
corresponds to the desired distribution, independent of the initialization.
We used a very conservative number of 500 steps. 

The MCMC has been set to contain 500 sampling points. It may be not
sufficient to have a smooth marginal PDF but more iterations are easy
to compute using the same algorithm using more computation time. For
the largest shape in the bivariate PDF, we used 5000 sampling points
instead. Let note the MCMC, sampling the final solution $\sigma_{M}(m)$
as: 
\begin{equation}
\left\{ \check{m}_{j}\right\} _{l},\:l=1,500
\end{equation}

The corresponding MCMC sampling of $\sigma_{D}(d)$ is estimated using:
\begin{equation}
\left\{ \check{r}_{i}\right\} _{l}=\left\{ F\left(\check{m}_{j},\Omega_{i}\right)\right\} _{l},\:l=1,500
\end{equation}

Please note that this method is somehow similar to genetic algorithms
\citep{Cord_PlanetBDRF_Icarus2003} and Particle Swarm Optimization
\citep{Beck_Photometryofmeteorites_I2012,Pommerol_PhotometricpropertiesMars_JGRP2013},
already used to solve this inverse problem empirically. Nevertheless,
those previous methods, although much faster, are only heuristic since
no convergence proof of the algorithms have been proposed. This is
not the case for the Metropolis rule, proposed here. 

The ability to rapidly find the ``best solution'' of heuristic method
is very convincing but these methods are not able to estimate the
uncertainties. In our case, the posterior PDF of a retrieved parameter
is not necessarily a gaussian distribution which is the main advantage
of the Bayesian method. Nevertheless, the Metropolis method can be
very slow, especially when the solution is well constrained (i.e.,
\emph{a posteriori} PDF with a very low standard deviation). 

In our case, all parameters have uniform physical prior distribution.
If the solution has an uncertainty of 10\% of the complete physical
domain in $M$ (for instance 0.5-0.6 for the parameter $b$, which
can vary from 0 to 1), the relevant subspace in the parameter space
is only $0.1^{6}=10^{-6}$ (for 6 parameters). It means that statistically,
only 1 iteration over $10^{6}$ is kept in the MCMC and all other
results are erased. To improve the computation time in this situation,
we propose to use a fast Monte Carlo method described in section \ref{sub:Fast-MCMC}.

\subsection{Description of the MCMC}

To describe the solution $\sigma_{M}(m)$, in addition to simple histograms,
additional statistical indicators on the MCMC are proposed, following
\citet{Fernando_SurfacereflectanceMars_JoGRP2013}:
\begin{itemize}
\item The average value $\hat{m}_{j}$ (mathematical expectation) of each
parameter $j$, and the estimated reflectance $\hat{r}_{i}$ at each
geometry $i$. 
\item The covariance matrix $\hat{C_{m}}$ in the model space can be estimated
from the MCMC. The $\hat{\sigma_{j}}$ standard deviation error bars
on each parameter $j$ are estimated from the covariance matrix elements
$\hat{C_{m}}_{jj}=\hat{\sigma_{j}^{2}}$.
\item The non-uniformity criterion $\hat{k}$. The parameters \emph{$m$}
are constrained if their marginal posterior PDF differs from the prior
state of information (i.e., uniform distribution in our case). In
order to distinguish if a given parameter is constrained we use the
non-uniformity criterion $\hat{k}$. \\
Central moments $\mu_{n}$ (such as the variance $\mu_{2}$ at order
two) are commonly used for statistical purpose while cumulants $k_{n}$
have the advantage to present unbiased statistical estimator for all
orders (\citep{Fisher1930}). The first four cumulants $k_{1},$ $k_{2}$,
$k_{3}$, $k_{4}$ of a uniform PDF are equal to $\frac{1}{2}$,
$\frac{1}{12}$, $0$, $-\frac{1}{120}$. We proposed the
parameter $\hat{k}$ to estimate the non-uniformity, defined as:
\begin{equation}
\hat{k}=\max\left|\frac{k_{1}-\frac{1}{2}}{\frac{1}{2}},\frac{k_{2}-\frac{1}{12}}{\frac{1}{12}},\frac{k_{3}}{\frac{1}{60}},\frac{k_{4}+\frac{1}{120}}{\frac{1}{120}}\right|
\end{equation}
We perform a numerical test of 10,000 uniform random vectors of 500
samples (identical to the MCMC) and compute $\hat{k}$ for each vectors.
Since the maximum is $\hat{k}$=0.47 for the most extreme vector,
we propose to consider non-uniform PDF for $\hat{k}>0.5$, which is
true with a probability higher than $99.99$ \%. For the inversion
purpose, since the a priori PDF on the parameters are uniform, if
the results of the inversion on one parameter has $\hat{k}<0.5$,
we conclude that this parameters is not constrained by the observations.
\end{itemize}

\subsection{Fast MCMC\label{sub:Fast-MCMC}}

The naive but accurate Monte Carlo Metropolis rule described in the
previous section may be not applicable in the case of long computation
time due to, either a large number of sampled geometries, or a small
data uncertainty. In the first case, the computation time is large
due to the time required to compute one direct model (one candidate
model). In the second case, the computation time is large because
the algorithm tests numerous non relevant parameter set solutions
(within the model space $M$ but far away from the actual solution). 

In order to speed up the computation time, it is possible to use an
adaptive Metropolis algorithm \citep{Haario_adaptiveMetropolisalgorithm_B2001}
but the solution (a posteriori PDF $\sigma_{M}(m)$) has to be close
to a gaussian distribution. This algorithm uses the last Markov Chain
to estimate a more relevant prior information $\hat{\rho_{M}}(m)=\mathcal{G}(\hat{m},\hat{C_{m}})$,
recursively closer to the actual solution. If the solution is gaussian,
the convergence of this method has been proven \citep{Haario_adaptiveMetropolisalgorithm_B2001}.
Also, this method is very convenient since it reduces very significantly
the number of steps. In our tests, the speedup can reach a factor
100 without significant differences, when the final solution is well
constrained. Table \ref{tab: MCMCvsfastMCMC} indicates an example
of results on a full BRDF (see fig. \ref{fig:Results_BRDF_geom9}
and \ref{fig:Results_BRDF_geom9invoptim}). Differences between estimated
parameters $\hat{m}$ for MCMC and fastMCMC are always much lower
than the estimated standard deviation $\hat{\sigma}$ so we can consider
as statistically equivalent.

\begin{table}
\begin{centering}
\begin{tabular}{|c|c|c|c|c|c|c|c|}
\hline 
 & algorithm & $b$ & $c$ & $\bar{\theta}$ & $\omega$ & $B_{0}$ & $h$\tabularnewline
\hline 
\hline 
$\hat{m}$ & MCMC & 0.45  & 0.45  & 7.42  & 0.89  & 0.50 & 0.48\tabularnewline
 & fastMCMC & 0.46 & 0.46  & 6.13  & 0.89  & 0.47  & 0.44\tabularnewline
\hline 
$\hat{\sigma}$  & MCMC & 0.06  & 0.08  & 4.30  & 0.02  & 0.28  & 0.29\tabularnewline
 & fastMCMC & 0.04  & 0.06  & 4.57  & 0.02  & 0.30  & 0.27\tabularnewline
\hline 
$\hat{k}$ & MCMC & 1.00  & 1.00  & 0.99  & 1.00  & 0.25  & 0.10\tabularnewline
 & fastMCMC & 1.00  & 1.00  & 1.01  & 1.00  & 0.25  & 0.44\tabularnewline
\hline 
\end{tabular}
\par\end{centering}

\protect\caption{Comparison between results of the retrieved parameters $\omega$,
$b$, $c$, $\bar{\theta}$, $B_{0}$ and $h$ using classical Metropolis
MCMC and fast MCMC, on average value $\hat{m}$, standard deviation
$\hat{\sigma}$ and non uniformity criteria $\hat{k}$. The data is
a full BRDF measurement at 96 geometrical conditions and using the
following parameter set: $b=0.5$, $c=0.5$, $\bar{\theta}=1^\circ$ ,
$\omega$=0.9, $h$=0, $B_{0}$=0, and 10\% uncertainties as described
in fig. \ref{fig:Results_BRDF_geom9} and \ref{fig:Results_BRDF_geom9invoptim}.
\label{tab: MCMCvsfastMCMC}. }
\end{table}

\section{Synthetic tests}

We perform several synthetic observations in different conditions,
to propagate the uncertainties from observations into the uncertainties
on the Hapke parameters. 

The first goal is to determine favorable geometric conditions to accurately
estimate the parameters for future spaceborne, in situ and laboratory
investigations (section \ref{sec:Uncertainties-propagation}). We
will study two difficult cases : EPF observation and one single image
/ disk resolved image. The case of a full BRDF is not relevant because
the uncertainties are small for all parameters $\omega$, $b$, $c$,
$\bar{\theta}$ with coherent values, as shown in the example in table
\ref{tab: MCMCvsfastMCMC}. The opposition effect parameters $B_{0}$
and $h$ can only be constrained for small phase angle measurements
and are out of the scope of this article. Those parameters are often
studied separately in the literature.

The second goal is to estimate if the Hapke hockey stick relationship
could be due to a non-linear effect on the inversion (section \ref{sec:Possible-origin-hockey-stick}).

For each test, we compute a perfect model $r_{i}$ in REFF (REFlectance
Factor) unit at known geometry $\Omega_{i}=(\theta_{0},\theta,g)_{i}$
using eq. \ref{eq:Hapke}, and known parameter $m_{j}$. The reflectance
in REFF unit is $REFF=\pi\cdot r(\theta_{0},\theta,g)/cos(\theta_{0})$.
We model the uncertainties on the measurement as a gaussian function,
independent from each geometry. The standard deviation level $\sigma_{i}$
at geometry $i$ is set to 10 \% of the observed reflectance $o_{i}$
in all the numerical tests, except when specified:

\begin{equation}
\sigma_{i}=\frac{o_{i}}{10}\label{eq:NoiseLevel}
\end{equation}
This value may be an upper limit for some spaceborne/laboratory instrumental
uncertainties but it shall give an upper limit of the final uncertainties
on the parameters. Also, taking all sources of error (including the
atmosphere correction), a noise level of 10\% is realistic \citep{Ceamanos_SurfacereflectanceMars_JoGRP2013,Fernando_SurfacereflectanceMars_JoGRP2013}.
For the case of CRISM data, the reflectance error at each geometry
were estimated at $\sigma_{i}=r_{i}/50$ for instance \citep{Fernando_SurfacereflectanceMars_JoGRP2013}.

\section{Uncertainties propagation\label{sec:Uncertainties-propagation}}

\subsection{Emission Phase Function (EPF)}

An Emission Phase Function (EPF) is a special configuration of observation
with one particular incidence direction (incidence angle $\theta_{0}$)
and a collection of emergence angle (emergence angles $\theta$) along
one single azimuthal plane. Table \ref{tab: angular configuration}
summarizes the different EPF conditions used. The EPF reflectance
set, usually represented as a function of the phase angle $g$, is
also called a photometric curve (Figure \ref{fig:Results_EPFexample_fit},
black curve). 

\begin{table}
\begin{centering}
\begin{tabular}{|c|c|}
\hline 
 & Angular configurations\tabularnewline
\hline 
\hline 
$\theta_{0}$ (deg.) & 30 ; 40 ; 50 ; 60 ; 70 ; 80\tabularnewline
\hline 
\hline 
$\theta$ (deg.) {[}case \#1{]} & {[}70, 50, 30, 10, -10, -30, -50, -70{]} \tabularnewline
\hline 
$\theta$ (deg.) {[}case \#2{]} & {[}70, 64, 58, 52, 47, 0, -47, -52, -58, -64, -70{]} \tabularnewline
\hline 
$\theta$ (deg.) {[}case \#3{]} & {[}80, 70, 60, 50, 40, 30, 20, 10, 0, -10, -20, -30, -40, -50, -60,
-70, -80{]} \tabularnewline
\hline 
\hline 
\{$\varphi_{1}$ ; $\varphi_{2}$ \} (deg.) & \{0 ; 180\} ; \{30 ;150\} ; \{60 ;120\} ; \{90 ; 90\}\tabularnewline
\hline 
\end{tabular}
\par\end{centering}

\protect\caption{Angular configurations used in the EPF synthetic tests.We tested:
(i) 6 incidence angles $\theta_{0}$, (ii) 3 sets of emission angle
$\theta$. The case \#1 represents a standard EPF, case\#2 corresponds
to a CRISM-like EPF observations and the case \#3 corresponds to a
favorable condition with a broad emission angle sampling typically
used in laboratory investigations, (iii) 4 sets of azimuth angle $\varphi$
($\varphi_{1}$: azimuth angle in the illumination direction, $\varphi_{2}$:
azimuth angle in the opposite illumination direction)\label{tab: angular configuration}}
\end{table}

\subsubsection{Uncertainties in one EPF example}

We simulate a standard EPF observations (Table \ref{tab: angular configuration},
case \#1) with a incidence angle $\theta_{0}=60^\circ$ along
the azimuthal plane \{$\varphi_{1}$ ; $\varphi_{2}$ \}=\{30; 210\}
resulting in a phase angle range from 29$^\circ$ to 122$^\circ$ ,
and using the following model parameter set: $\omega=0.9$, $b=0.8$,
$c=0.1$, $\bar{\theta}=15^\circ$, $B_{0}=0$ and $h=0$
corresponding to a bright material with a narrow forward scattering
behavior and rough topographic surface. This surface corresponds to
granular soil with small grain size and round shape. A similar set
of parameters have been observed in the laboratory measurements of
olivine at 700 nm \citep{Souchon_experimentalstudyof_I2011}. Then,
we invert the synthetic dataset with an uncertainty level set at 10\%
of the reflectance (one standard deviation). We examine the final
solution estimated from the last 500 iterations of the Markov Chain.
One have to remind that the discrepancies between the solution and
the initial data are not due to the retrieval method itself, but by
the lack of information in the available data (poor geometric sampling,
large uncertainties).

\begin{figure}
\includegraphics[clip,height=0.6\textwidth]{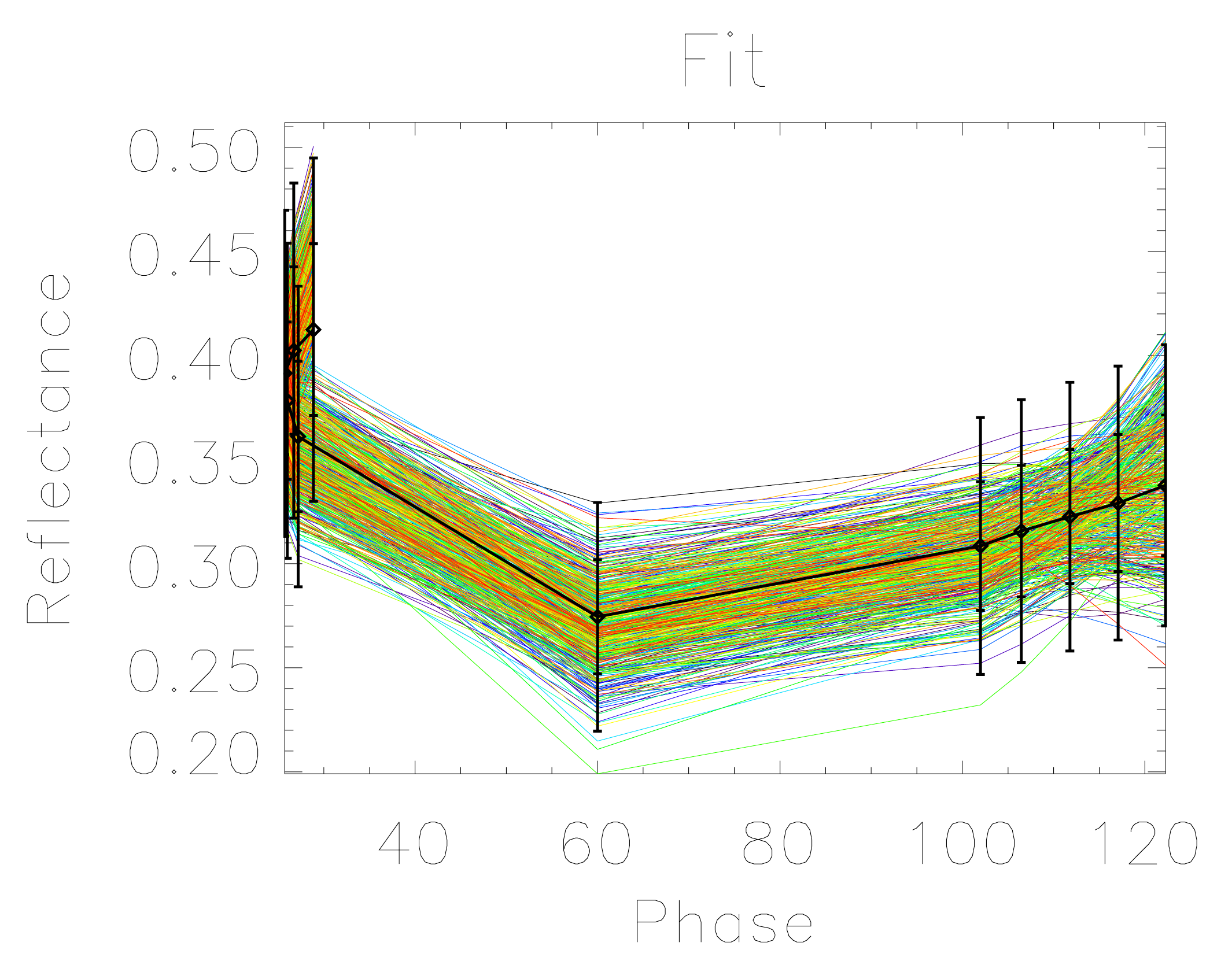}\centering\protect\caption{Example of a result on a synthetic observation with 10\% uncertainty.
The black curve represents the synthetic data with one and two standard
deviations. Light color curves represent 500 sampled solutions from
the Monte Carlo Markov Chain.}
\label{fig:Results_EPFexample_fit} 
\end{figure}

Figure \ref{fig:Results_EPFexample_fit} presents the synthetic data
and the fit of the 500 sampled solutions. The solutions are fitting
the synthetic data with the expected tolerance (95\% of the fits inside
the 2 $\sigma$ data error). 

\begin{figure}
\includegraphics[clip,width=0.9\textwidth]{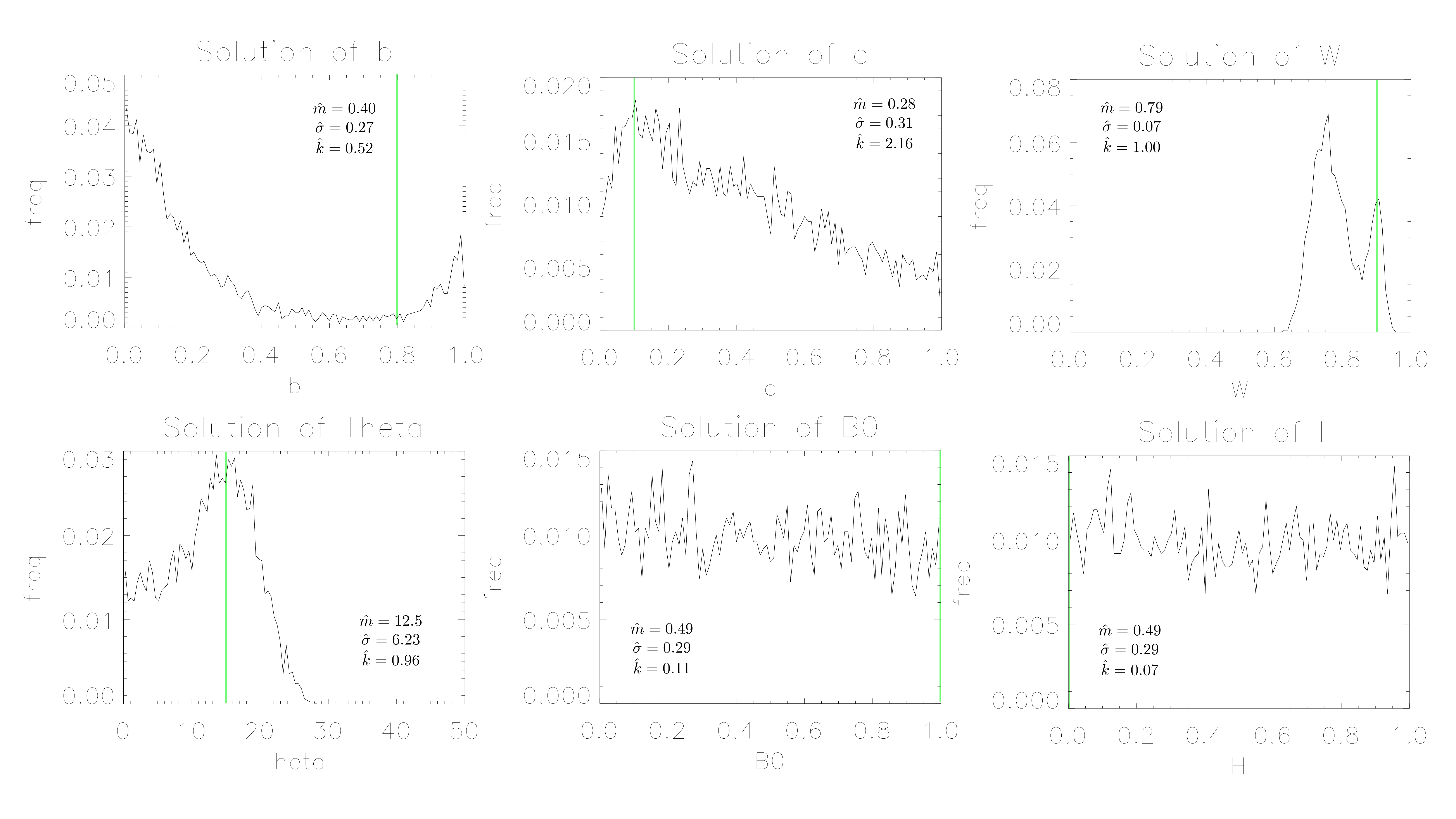}\centering\protect\caption{Probability Density Function (PDF) of the solution, for each parameter
of the Hapke model on the same example as fig. \ref{fig:Results_EPFexample_fit}.
Each plot represents the histogram of the 500 acceptable solution
from the Monte Carlo Markov Chain. The color line represent the initial
parameters set.}
\label{fig:Results_EPFexample_histSingle} 
\end{figure}

Figure \ref{fig:Results_EPFexample_histSingle} shows the histogram
of the Hapke parameters estimated from the 500 sampled solutions.
The opposition effect parameters $B0$ and $h$ present a flat histogram
and have a non-uniformity criterion $\hat{k}<0.5$, suggesting that
both parameters are not constrained. These results are directly related
to the lack of observations at phase angle lower than 20$^\circ$
to observe the opposition effect. The single scattering albedo histogram
shows a double peak around the expected value ($\omega$=0.9), showing
the effect of the Hapke model non linearity. 

This example shows that the Bayesian inversion is able to identify
several possible solutions due to data uncertainty and/or geometric
sampling. The apparition of the double peaks on the parameter $\omega$
has been identified to chase the limitations of geometric diversity
and/or reflectance uncertainty \citep{Fernando_SurfacereflectanceMars_JoGRP2013}.
Since, usually $\omega$ is the best-constrained parameter in photometric
modeling, the double peak in $\omega$ is also an indicator of low-constrain
on other Hapke parameters. The parameters $c$ and $\bar{\theta}$
show a broad PDF with a maximum close to the expected solution. Interestingly,
the parameter $b$ has a PDF that is not peaking to the expected solution,
most probably because the phase range 29$^\circ$ -122$^\circ$
at 10\% uncertainties is not sufficient to constrain it.

\begin{figure}
\includegraphics[clip,height=0.9\textwidth]{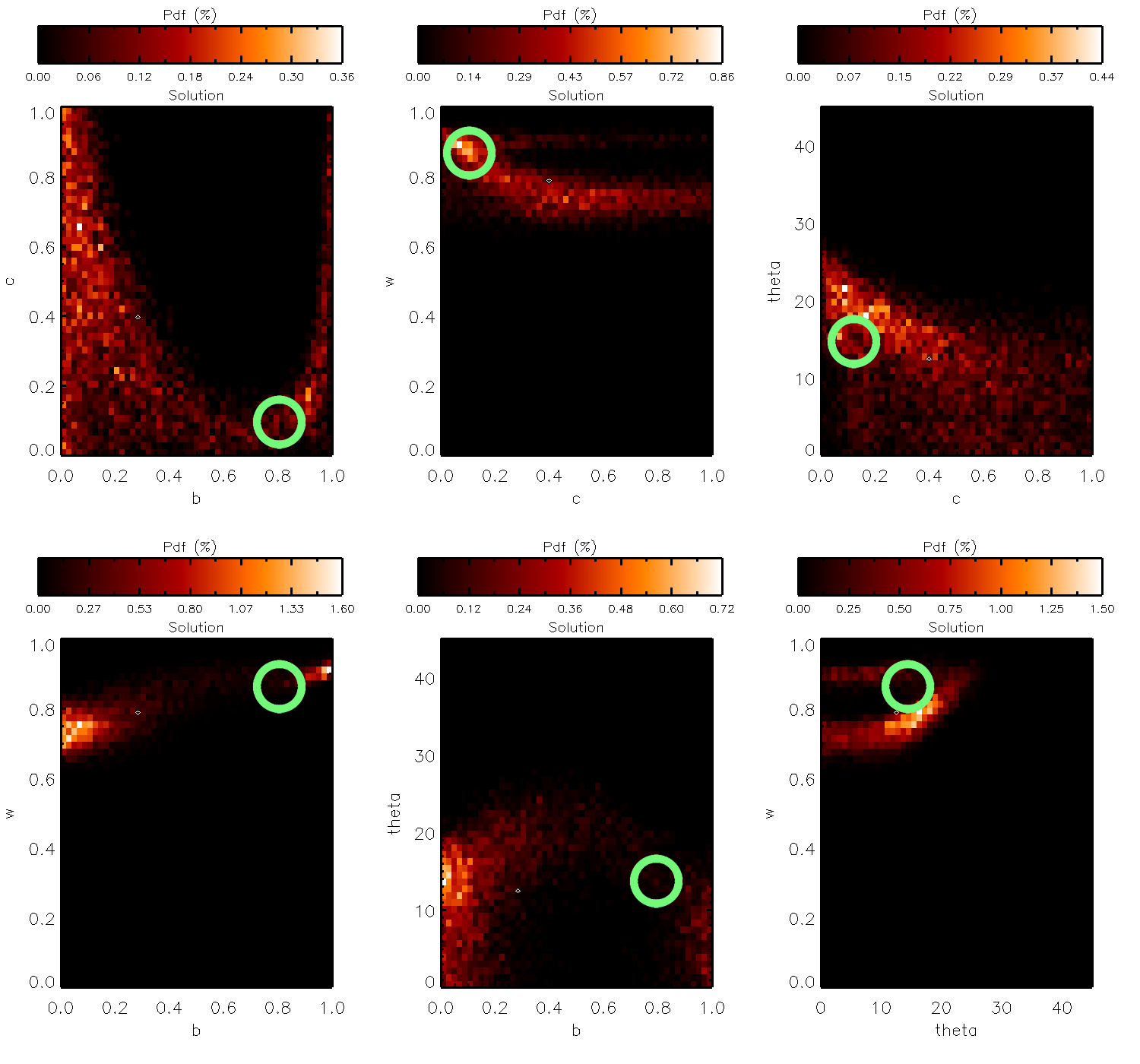}\centering\protect\caption{Probability Density Function (PDF) of the solution, for each couple
of constrained parameters ($\omega$, $b$, $c$, $\bar{\theta}$)
on the same example as fig. \ref{fig:Results_EPFexample_fit} and
\ref{fig:Results_EPFexample_histSingle}. The black/white diamonds
represent the average of the PDF. The green circles represent the
expected values for each parameter. }
\label{fig:Results_EPFexample_histDouble} 
\end{figure}

Figure \ref{fig:Results_EPFexample_histDouble} presents the bivariate
histogram for couples of parameters, permitting to study the combined
effects of two parameters. It is clearly demonstrated that all parameters
are correlated. For instance, the $b$ vs $c$ histogram clearly shows
a ``U'' shape covering a large part of the model space $M$, but
with a strong correlation, excluding medium $b$ and strong $c$ solutions.
The single scattering albedo $\omega$ is better estimated with low
$c$ values, but higher $c$ values are compensated by slightly lower
$\omega$, demonstrating the complex correlation of $\omega$ and
$c$.

This test shows that the Bayesian inversion is able to sample complex
solutions in model parameters $M$ which is not the case for classic
inversion procedures based on minimization techniques. It also demonstrates
that the maximum likelihood may be a wrong indicator of the whole
solution and that uncertainties may have a very complex shape due
to correlation between parameters.

\subsubsection{Favorable geometric sampling and uncertainties\label{sub:Favorable-geometric-sampling}}

The objective of this sub-section is to evaluate the favorable geometric
conditions in order to better estimate the Hapke photometric parameters
by simulating various surface scattering properties. These tests allow
to evaluate the influence of the EPF geometric sampling in the uncertainties
of the retrieved model parameters. These synthetic results should
give constraints for current spaceborne data analyses (e.g., CRISM/MRO,
HRSC/MRO, OMEGA/MEx for Mars, VIMS/Cassini for Titan, VIRTIS/ Rosetta
for 67P/Churyumov-Gerasimenko, LROC/LRO for the Moon, etc) and future
instrumental and laboratory investigations. The following tests are
done only for realistic soil cases, with $b$/$c$ in the Hapke hockey-stick
relationship. The opposition effect is not studied here because it
is usually investigated separately from the global photometric data,
this effect being only rarely observable in spaceborne data.

We perform synthetic tests, using various surface scattering properties
in order to cover the range of properties which can be observed in
natural environments (i.e., different Hapke parameter sets):
\begin{itemize}
\item 2 single scattering albedos ($\omega$): 0.3; 0.9
\item 6 parameters ($b$ and $c$ couples) taken along the hockey stick
relation: L1($b$=0.3/$c$=1.0), L2(0.3/0.8), L3(0.3/0.5), L4(0.5/0.2),
L5(0.8/0.1), L6(0.8/0.1)
\item 2 macroscopic roughnesses ($\bar{\theta}$): 0$^\circ$ ; 15$^\circ$
\item Opposition effect parameters are set to ignore this effect: $B_{0}$=0
and $h$=0.
\end{itemize}
We perform the synthetic tests under various geometric configurations
summarized in Table \ref{tab: angular configuration}: 6 incidence
angles, 4 azimuthal modes and 2 cases of emission angles samplings
where the case \#2 to a CRISM-like EPF observation with a poor-sampling
and a narrow emission angle range (11 configurations) and the case
\#3 corresponds to a EPF with a well-sampling and broad emergence
angle range (17 configurations). In what follows, we focus on the
results of the case \#2 as for example but all the results of the
case \#3 are presented in the Supplementary Materials.

All 576 configurations are summarized in a single graph by computing
the difference between initial parameter value of the synthetic test
$m_{j}$ and the estimated one from the Bayesian inversion $\check{m}_{j}$
(with $j$ the index of parameter and $l$ the index of the Markov
Chain):

\begin{equation}
\left\{ \triangle m_{j}\right\} _{l}=\left\{ \check{m}_{j}\right\} _{l}-m_{j}\label{eq:DefinitionDeviation}
\end{equation}

This quantity represents the deviation between the initial ``true''
parameters and the state of information present in the EPF data, given
the geometric sampling and uncertainties. The histogram of $\left\{ \triangle m_{j}\right\} _{l}$
represents the PDF of uncertainties. By tracing marginal PDF, it is
possible to chase the effect of each parameter on the uncertainties.
One has to remind that the deviation between the solution and the
initial value are not due to the retrieval method itself, but to the
lack of information in the available dataset.

\begin{figure}
\includegraphics[clip,width=0.8\textwidth]{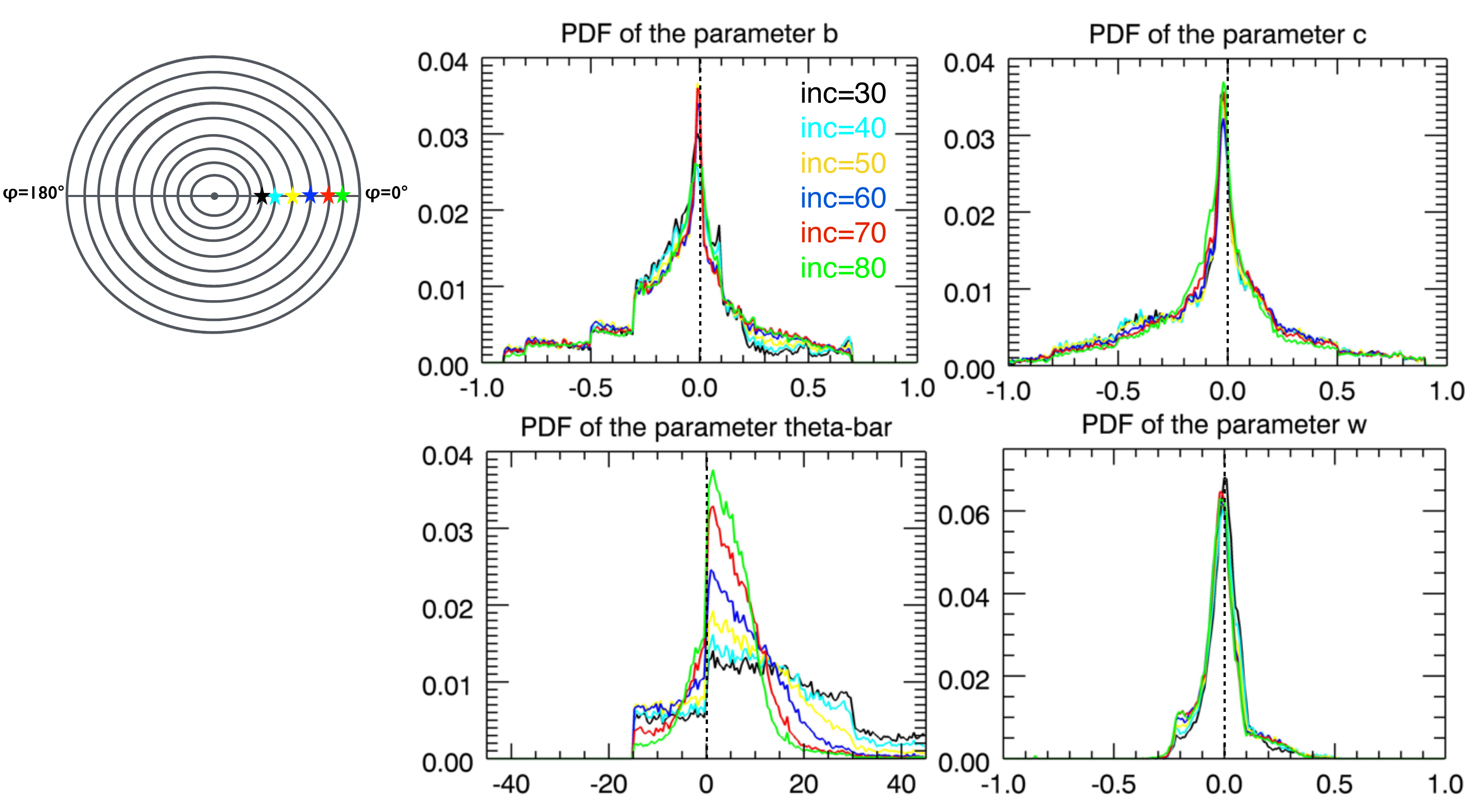}\centering\protect\caption{Difference between true parameter and the Bayesian solution (see eq.
\ref{eq:DefinitionDeviation}), for 6 incidence: 30$^\circ$ , 40$^\circ$ ,
50$^\circ$ , 60$^\circ$ , 70$^\circ$ , 80$^\circ$ . All
other parameters are included.}
\label{fig:Results_bias_incidence} 
\end{figure}

\begin{figure}
\includegraphics[clip,width=0.8\textwidth]{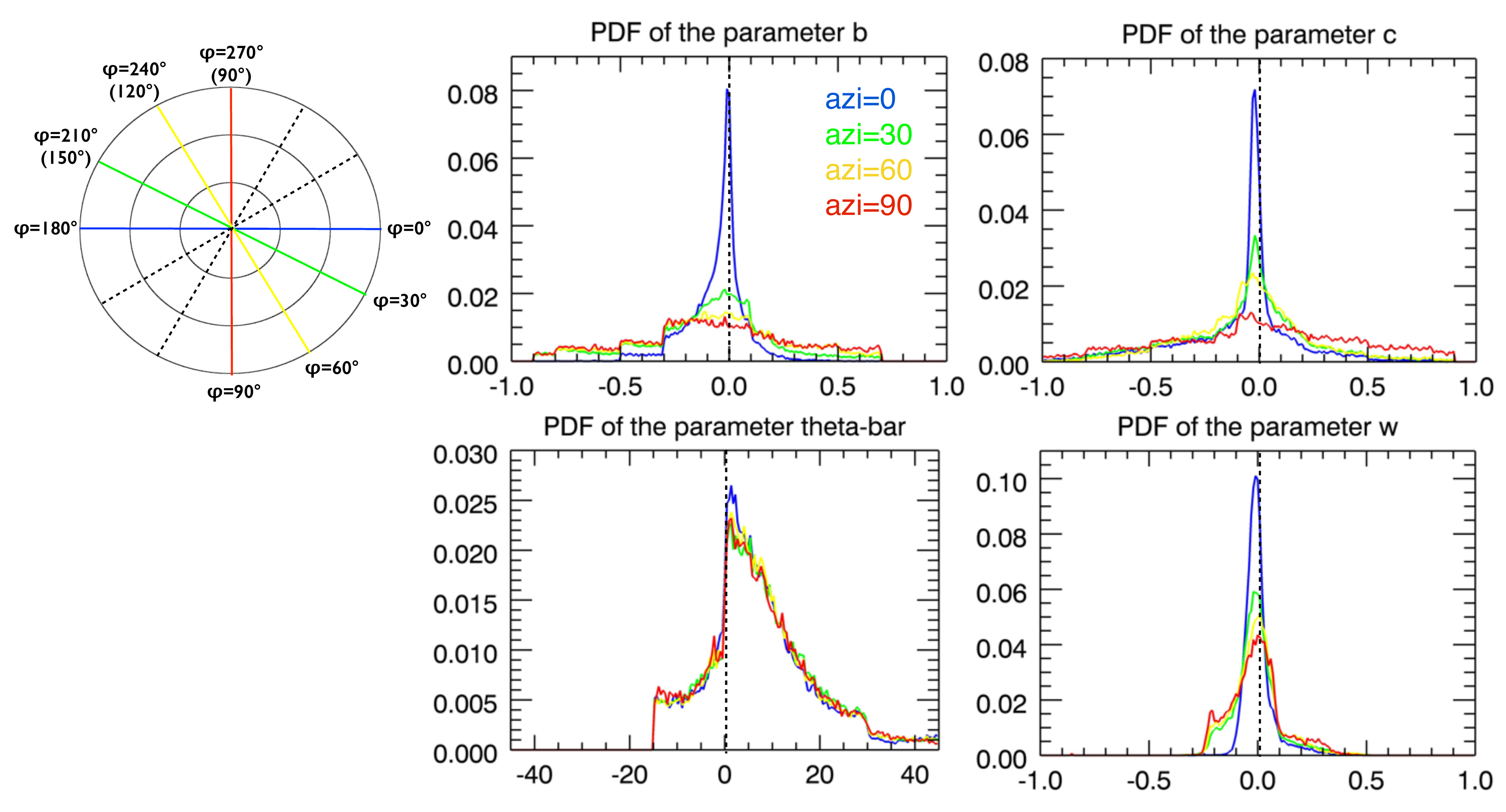}\centering\protect\caption{Difference between true parameter and the Bayesian solution (see eq.
\ref{eq:DefinitionDeviation}), for 4 azimuth: 0$^\circ$ , 30$^\circ$ ,
60$^\circ$ , 90$^\circ$ . All other parameters are included.}
\label{fig:Results_bias_azimuth} 
\end{figure}

\begin{figure}
\includegraphics[clip,width=0.8\textwidth]{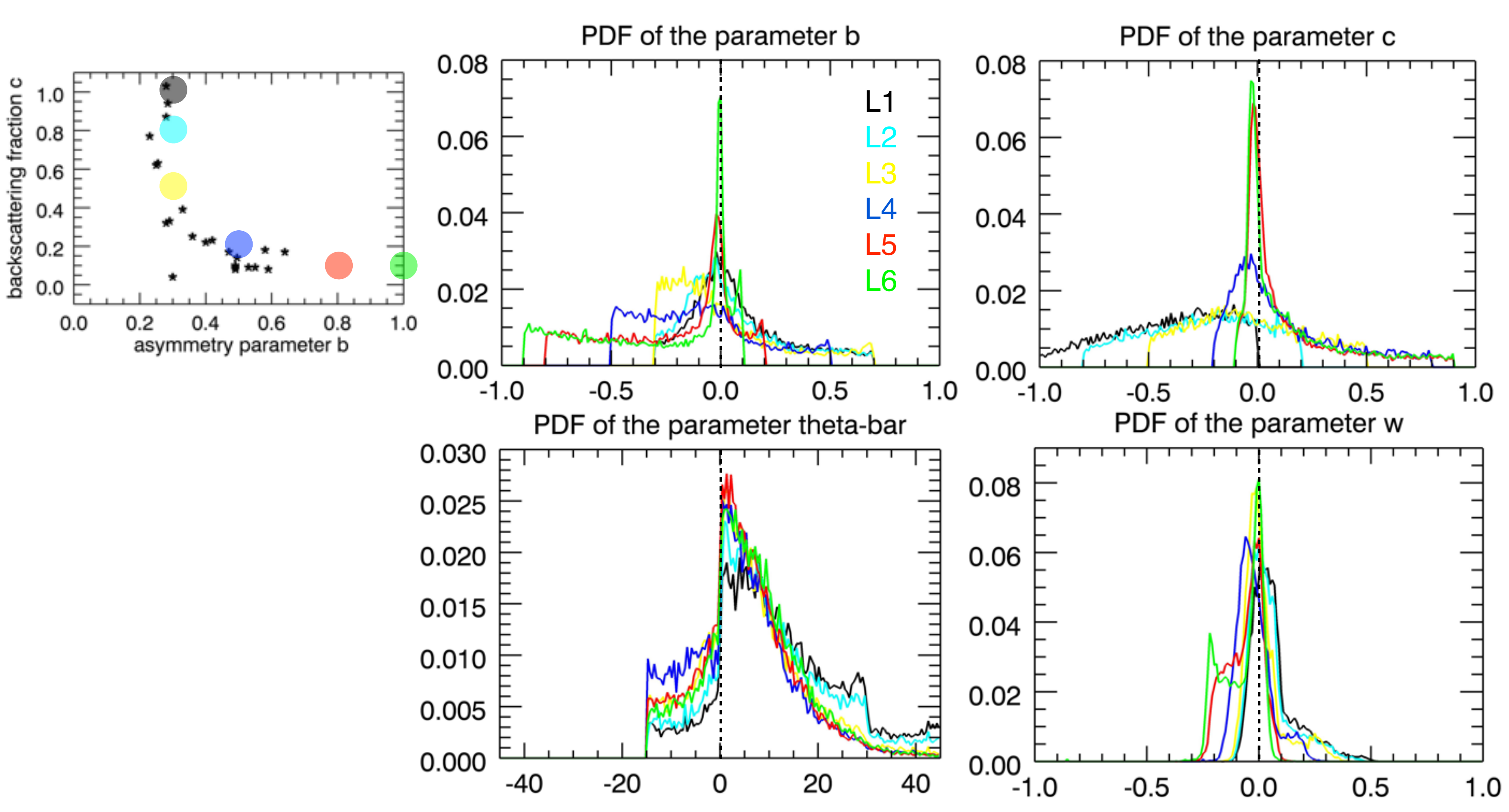}\centering\protect\caption{Difference between true parameter and the Bayesian solution (see eq.
\ref{eq:DefinitionDeviation}), for 6 configurations of phase function
parameters in the hockey stick: \#1($b$=0.3/$c$=1.0), \#2(0.3/0.8),
\#3(0.3/0.5), \#4(0.5/0.2), \#5(0.8/0.1), \#6(0.8/1.0) All other parameters
are included.}
\label{fig:Results_bias_L} 
\end{figure}

\begin{figure}
\includegraphics[clip,width=0.8\textwidth]{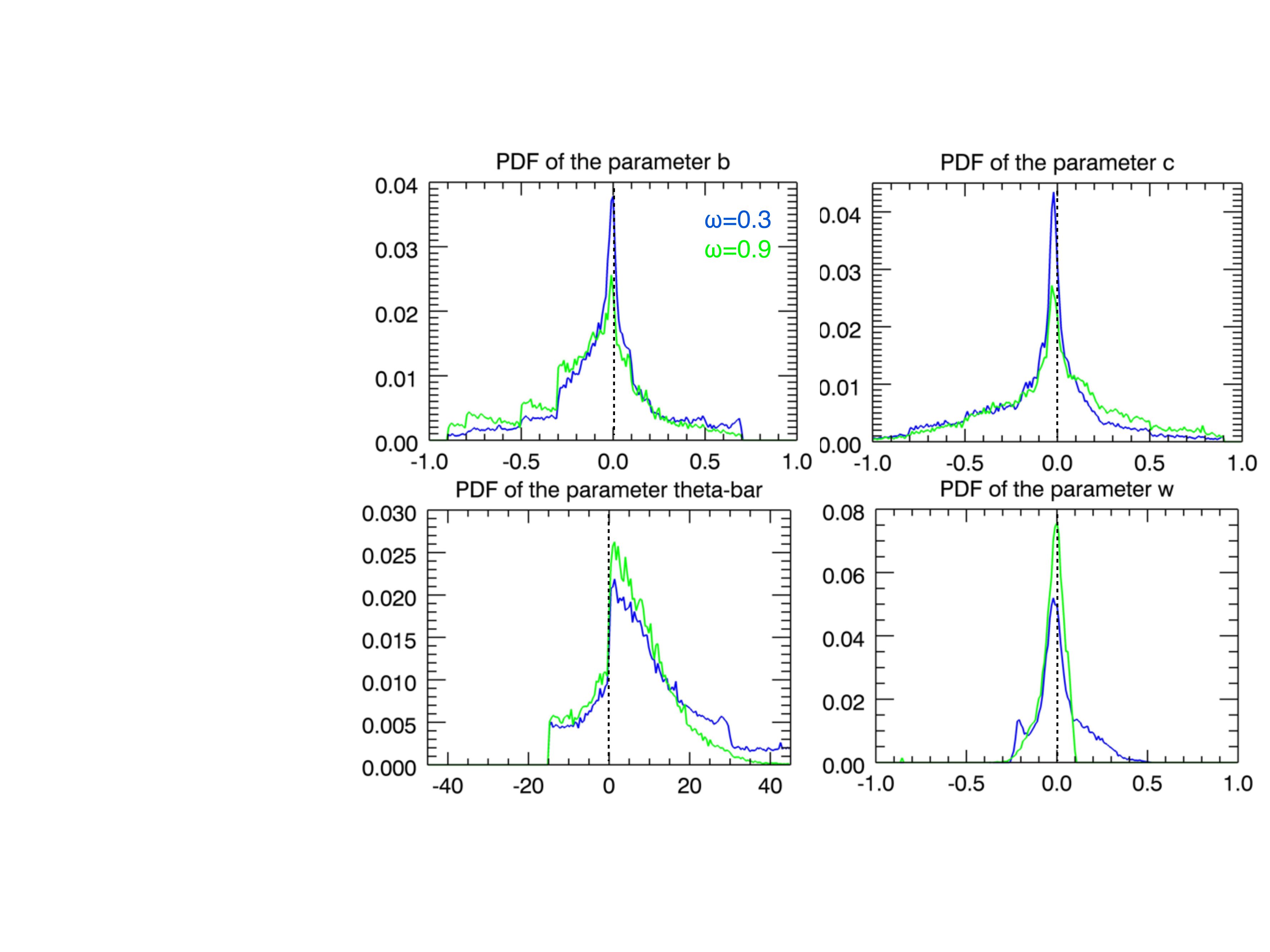}\centering\protect\caption{Difference between true parameter and the Bayesian solution (see eq.
\ref{eq:DefinitionDeviation}), for 2 single scattering albedo $\omega$:
0.3, 0.9. All other parameters are included.}
\label{fig:Results_bias_w} 
\end{figure}

\begin{figure}
\includegraphics[clip,width=0.8\textwidth]{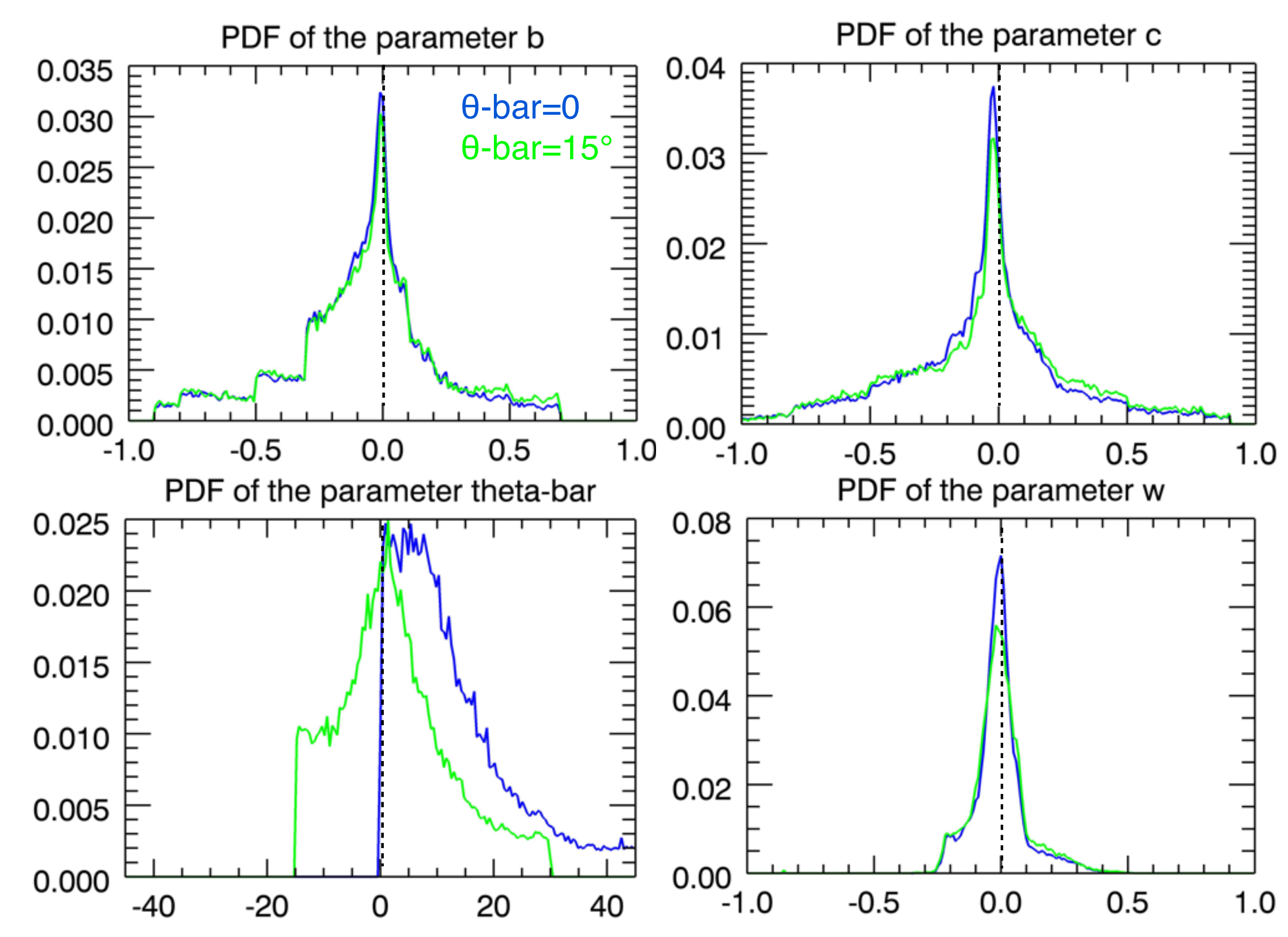}\centering\protect\caption{Difference between true parameter and the Bayesian solution (see eq.
\ref{eq:DefinitionDeviation}), for 2 macroscopic roughness $\bar{\theta}$:
0$^\circ$ , 15$^\circ$ All other parameters are included.}
\label{fig:Results_bias_roughness} 
\end{figure}

Figures \ref{fig:Results_bias_incidence} to \ref{fig:Results_bias_roughness}
present the deviation for one studied parameter each: geometric configurations:
incidence angles (fig. \ref{fig:Results_bias_incidence}), azimuthal
modes (fig. \ref{fig:Results_bias_azimuth}), model parameter: $c$
and $b$ couple (L1 to L6) (fig. \ref{fig:Results_bias_L}), $\omega$
(fig. \ref{fig:Results_bias_w}) and $\bar{\theta}$ (fig. \ref{fig:Results_bias_roughness})
by averaging the effects of all other parameters. The results show
that the distributions of $\left\{ \triangle m_{j}\right\} _{l}$
are always maximum on 0, showing that the maximum likelihood is globally
unbiased. It demonstrates that a single EPF is enough to estimate
accurate model parameters in most of the cases.

Nevertheless, the pick and the tails shape of the PDF of uncertainties
are not identical for all marginal probabilities, showing that the
CRISM-like EPF sampling can be optimized for some geometries, or for
some surface parameters.

One has to note that the roughness parameter is tested for the case
$\bar{\theta}=0$. Since $\bar{\theta\geq0}$ by definition, the deviation
cannot be symmetrical ($\left\{ \triangle m_{j}\right\} _{l}\geq0$),
this significantly biases the deviation PDF. 

In detail, we can take the following conclusions:
\begin{itemize}
\item The single scattering albedo parameter $\omega$ presents the narrower
deviation PDF in comparison to other Hapke parameters (fig. \ref{fig:Results_bias_incidence}
to \ref{fig:Results_bias_roughness} ). Thus $\omega$ is the best
constrained parameter on EPF measurement.
\item The influence of the geometric sampling on the deviation is very important:
both incidence angle (fig. \ref{fig:Results_bias_incidence}) and
azimuthal mode (fig. \ref{fig:Results_bias_azimuth}) significantly
change the global uncertainties:

\begin{itemize}
\item Incidence has a strong effect on the estimation of surface roughness:
the larger the incidence, the better the estimation of $\bar{\theta}$.
\item Azimuth has a strong effect on $b$, $c$ and $\omega$: the closer
to principal plane, the better the estimation. This effect is also
present on $\bar{\theta}$ but more moderately.
\item We can interpret this behavior by the diversity of phase angles necessary
to constraint the photometric parameters, both at low ($g<30{^\circ}$)
and high ($g>100{^\circ}$) ranges. Observations acquired at high
incidence angles and/or close to the principal plane allow to have
the broadest phase angle ranges corresponding to the most favorable
conditions. A departure from principal plane as low as 30$^\circ$
leads to significant increase of the uncertainties.
\end{itemize}
\item The influence of the model parameter values on deviation play a second
order role: 

\begin{itemize}
\item The phase function parameters seem to have a strong influence on the
retrieved parameters. In particular, the parameters $b$, $c$ and
$\omega$ are significantly better estimated for surface materials
with a narrow forward behavior (e.g., L6) (fig. \ref{fig:Results_bias_L}). 
\item The single scattering albedo $\omega$ has a moderate effect on the
estimation of the parameters $c$ and $\omega$: (i) the estimation
of parameter $\omega$ is better when the albedo is high, (ii) the
estimation of the parameter $c$ is better when the albedo is lower
(fig. \ref{fig:Results_bias_w}). 
\item The macroscopic roughness $\bar{\theta}$ has a small effect on the
estimation of the parameters $b$, $c$ and $\omega$: the parameter
estimations are better when the parameter $\bar{\theta}$ is lower.
(fig. \ref{fig:Results_bias_roughness}).
\end{itemize}
\item Those conclusions are also valid for the case of favorable EPF (case
\#3, presented in the Supplementary Materials), except that the uncertainties
are slightly reduced in this case. 
\end{itemize}
A similar study has been proposed on one experimental dataset using
the genetic algorithm \citep{Souchon_experimentalstudyof_I2011}.
The authors recommended a ``regular coverage of the bidirectional
space in incidence, emission, azimuth, and consequently phase angles''
and showed that ``reliable photometric estimates can be produced
with a limited set of angular configurations (on the order of a few
tens)''. Our conclusions demonstrate that this experimental work
can be extended even in the principal plane when large phase angle
range are available.

\subsection{One single observation of a rough surface / disk resolved image \label{sub:One-single-observation}}

One single image of a known rough surface has very limited angle geometry.
Such observational condition corresponds to the case where each pixel
of the image is assimilated to a facet with known orientation. We
assume that the surface properties are spatially homogeneous over
a large amount of pixel in order to estimate the surface properties
by combining several adjacent pixels. This case is equivalent to a
disk resolved measurement of a planetary body assumed to be homogeneous
in surface properties, at one single phase image. We study this very
difficult condition, in order to estimate if it is possible to constrain
the photometric parameters. 

For each facet (or equivalently for each image pixel), the incident
and emergent rays are parallel, with the same phase angle. However,
since each facet has its own orientation, there is a variability of
local incidence, emergence and azimuth angles. We propose to use a
typical observation condition of incidence $\theta_{0}=40^\circ$
, emergence $\theta=0^\circ$ on 24 facets with 6 slopes
$\theta_{s}=5^\circ$ , 10$^\circ$ , 20$^\circ$ ,
30$^\circ$ , 40$^\circ$ , 50$^\circ$ and 4 azimuth of slope
$\phi_{s}=0^\circ$, 90$^\circ$ , 180$^\circ$ ,
270$^\circ$ . Each facet is thus defined by a couple of slope and
azimuth angles $(\theta_{s},\phi_{s})$. The phase angle is always
$g=40{^\circ}$, but the local incidence $\tilde{\theta_{0}}$ on
the facet varies from 0$^\circ$ to 90$^\circ$ , the local
emergence $\tilde{\theta}$ from 5$^\circ$ to 50$^\circ$
and the local azimuth $\tilde{\phi}$ from 0$^\circ$ to 180$^\circ$ .
The relationship between slopes, azimuth of slopes and local incidence
and emergence are the following \citet{Hapke_Book1993}:
\begin{equation}
\cos\tilde{\theta_{0}}=\cos\theta_{0}\cos\theta_{s}+\sin\theta_{0}\sin\theta_{s}\sin\phi_{s}
\end{equation}

\begin{equation}
\cos\tilde{\theta}=\cos\theta\cos\theta_{s}+\sin\theta\sin\theta_{s}\sin\phi_{s}
\end{equation}

We test $\omega=0.9$, $\bar{\theta}=1^\circ$ and three
values of the $b$/$c$ in the hockey stick : \#1($b$=0.3/$c$=1.0),
\#4(0.5/0.2), \#6(1.0/0.1).

\begin{figure}
\includegraphics[clip,height=0.9\textwidth]{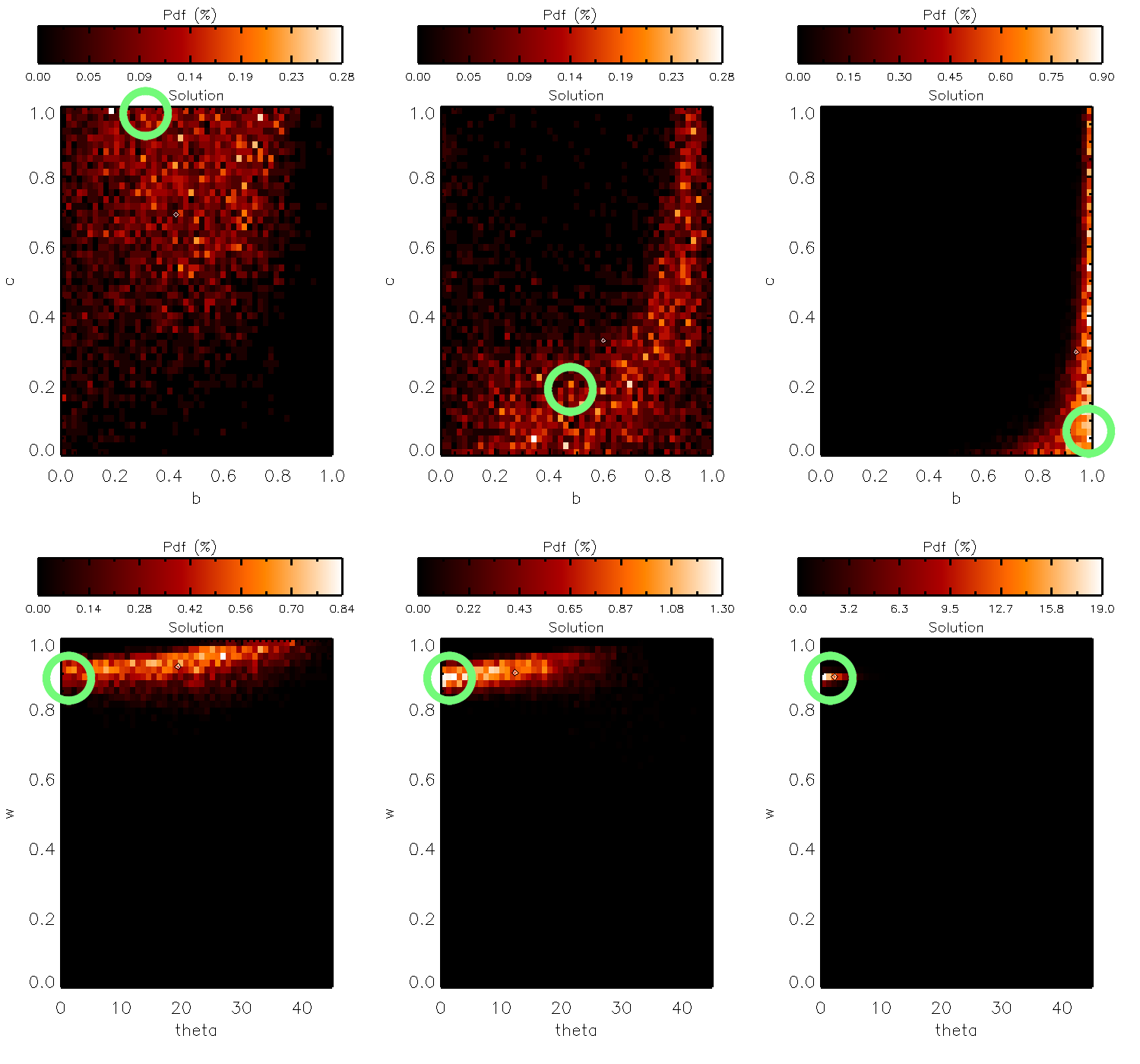}\centering\protect\caption{Probability Density Function of the solution, for each couple of constrained
parameters ($\omega$, $b$, $c$, $\bar{\theta}$) derived from the
simulation of a single observation of a known rough surface with 24
facets (see the text for more detail). At the left case \#1($b$=0.3/$c$=1.0).
In the middle, case \#4(0.5/0.2). At right case \#6(1.0/0.1). The
black/white diamonds represent the average calculated from of the
PDF. The green circles represent the expected values for each parameter. }
\label{fig:Results_OneSingleObservation_histDouble} 
\end{figure}

The results are plotted in figure \ref{fig:Results_OneSingleObservation_histDouble},
in terms of bi-variate histograms describing $b$ vs $c$, and $\omega$
vs $\bar{\theta}$. The parameter $\omega$ seems to be well constrained
in all cases but the parameter $\bar{\theta}$ is only constrained
in the case of a strong narrow forward scattering. The particulate
phase function parameters $b$ is only constrained in the extreme
cases of the hockey stick (cases \#1 and \#6). The parameter $c$
is not constrained.

These results indicate that information of the surface properties
can be retrieved even in the case of one single observation with known
geometry for each facet. Especially $\omega$ can be retrieved with
small uncertainties (1$\sigma$ uncertainties $\leq0.1$ ), but also
$b$, $c$ and in some extent $\bar{\theta}$, in the case of very
extreme phase function.

\section{Possible origin of the hockey stick relationship\label{sec:Possible-origin-hockey-stick}}

We exhibit here particular conditions of Emission Phase Function (EPF)
and Bidirectional Reflectance Distribution Function (BRDF) that could
be a possible bias at the origin of the hockey stick relationship,
so we focus in this section on the parameters $b$ and $c$. We set
$\omega$=0.9, $\bar{\theta}=1^\circ$ , $h$=0, $B_{0}$=0 and
data uncertainties at a level of 10\% as previously defined. The results
are expected to be only weakly dependent of $\omega$ and $\bar{\theta}$
(see section \ref{sub:Favorable-geometric-sampling}). These surface
properties correspond to granular soil with small grain size. Similar
parameters have been observed in sulfate terrain on Mars \citep{Johnson_Opportunity_JGR2006},
and in various samples \citep{Souchon_experimentalstudyof_I2011},
such basalt including feldspar grains, pyroclastic grains and olivine
grains.

We test here all configurations of $b$/$c$ covering the whole parameter
spaces from 0 to 1 and see if the resulting uncertainties can bias
the interpretation or not. We insist on the fact that materials with
both $b$$\geq$ 0.5 and $c$ $\geq$ 0.5 have never been observed.
Nevertheless, we tested these configurations in order to study the
potential bias when analyzing such case.

\subsection{Principal plane EPF\label{sub:Principal-plane-EPF}}

\begin{figure}
\includegraphics[bb=0bp 0bp 1500bp 2100bp,clip,height=0.9\textheight]{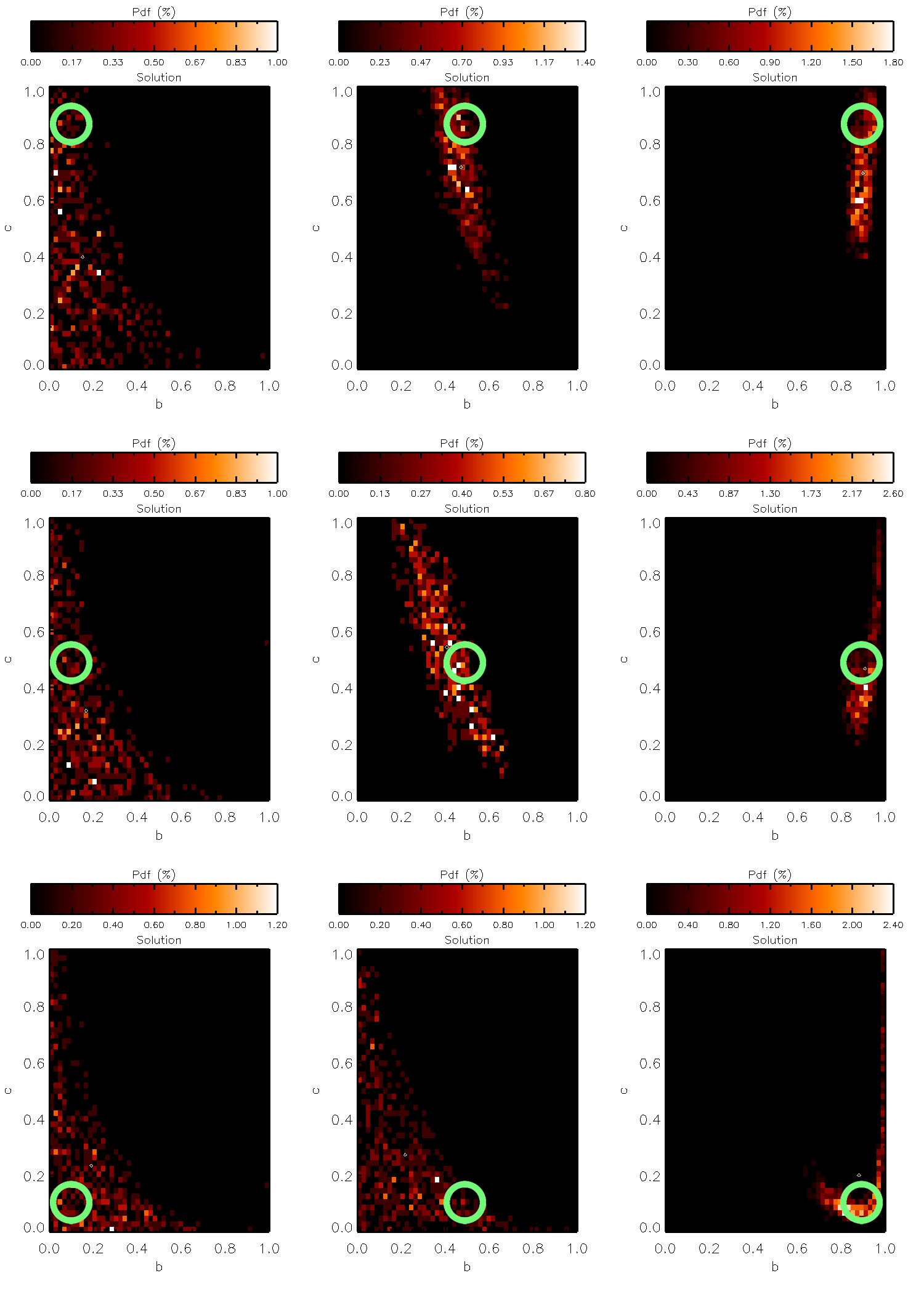}\centering\protect\caption{Probability density function of the $b$ and $c$ parameters using
a standard principal plane EPF observation (8 emergence configurations:
70$^\circ$ , 50$^\circ$ , 30$^\circ$ , 10$^\circ$ , -10$^\circ$ ,
-30$^\circ$ , -50$^\circ$ , -70$^\circ$ ), incidence=40$^\circ$ ,
azimuth=$\left\{ 0,180^\circ\right\} $ and using the
following parameter set: $\omega$=0.9, $\bar{\theta}=1^\circ$ ,
$h$=0, $B_{0}$=0, data uncertainties 10\% and 9 combinations of
$b$=0.1/0.5/0.9 and $c$=0.1/0.5/0.9. The black/white diamonds represent
the average of the PDF. The green circles represent the expected values
of the parameters $b$ and $c$. }
\label{fig:Results_EPF-CRISM_Az0_geom1} 
\end{figure}

Figure \ref{fig:Results_EPF-CRISM_Az0_geom1} presents the results
for a standard principal plane EPF observation (Table \ref{tab: angular configuration}
case \#1) (i.e. poor-sampling of emission angles, mostly the case
for spaceborne instruments like CRISM). It shows that low $b$ ($<0.5$)
imply poor constraint on $c$ and very large uncertainties on $b$.
For $b$$\geq$ 0.5 and $c$ $\geq$ 0.5, $b$ has small uncertainties
and $c$ has medium uncertainties.

\begin{figure}
\includegraphics[bb=0bp 0bp 1500bp 2100bp,clip,height=0.9\textheight]{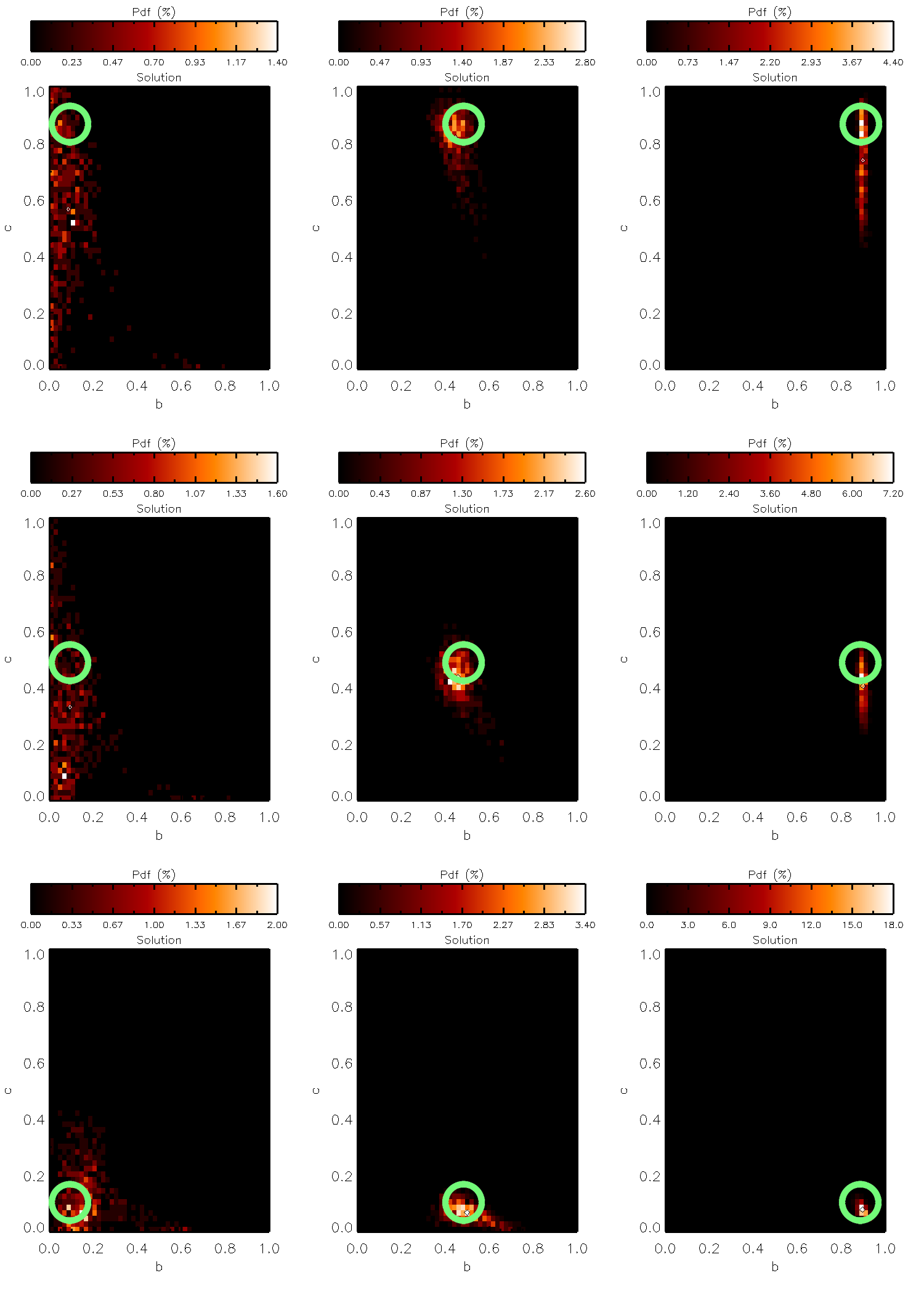}\centering\protect\caption{Probability density function of the $b$ and $c$ parameters using
a favorable principal plane EPF observation (17 emergence configurations:
80$^\circ$ , 70$^\circ$ , 60$^\circ$ , 50$^\circ$ , 40$^\circ$ ,
30$^\circ$ , 20$^\circ$ , 10$^\circ$ , 0$^\circ$ , -10$^\circ$ ,
-20$^\circ$ , -30$^\circ$ , -40$^\circ$ , -50$^\circ$ ,
-60$^\circ$ , -70$^\circ$ , -80$^\circ$ ), incidence=75$^\circ$ ,
azimuth=$\left\{ 0,180^\circ\right\} $ and using the
following parameter set: $\omega$=0.9, $\bar{\theta}$=1$^\circ$ ,
$h$=0, $B_{0}$=0, uncertainties 10\% and 9 combinations of $b$=0.1/0.5/0.9
and $c$=0.1/0.5/0.9. The black/white diamonds represent the average
of the PDF. The green circles represent the expected values of the
parameters $b$ and $c$. }
\label{fig:Results_EPF_Az0_geom2} 
\end{figure}

Figure \ref{fig:Results_EPF_Az0_geom2} presents the results for a
favorable principal plane EPF observation (i.e., well sampling of
emission angles which can be obtained at laboratory, on in situ in
very favorable conditions). The solution is clearly better constrained
than for the standard principal plane (see fig. \ref{fig:Results_EPF-CRISM_Az0_geom1}).
It demonstrates that even at 10\% data uncertainties, the correct
behavior (backward/forward and narrow/broad) can be retrieved from
a single EPF observation if the observation is taken in or close to
the principal plane.

\subsection{Effect of azimuth on EPF\label{sub:Effect-of-azimuth}}

\begin{figure}
\includegraphics[bb=0bp 0bp 1500bp 2100bp,clip,height=0.9\textheight]{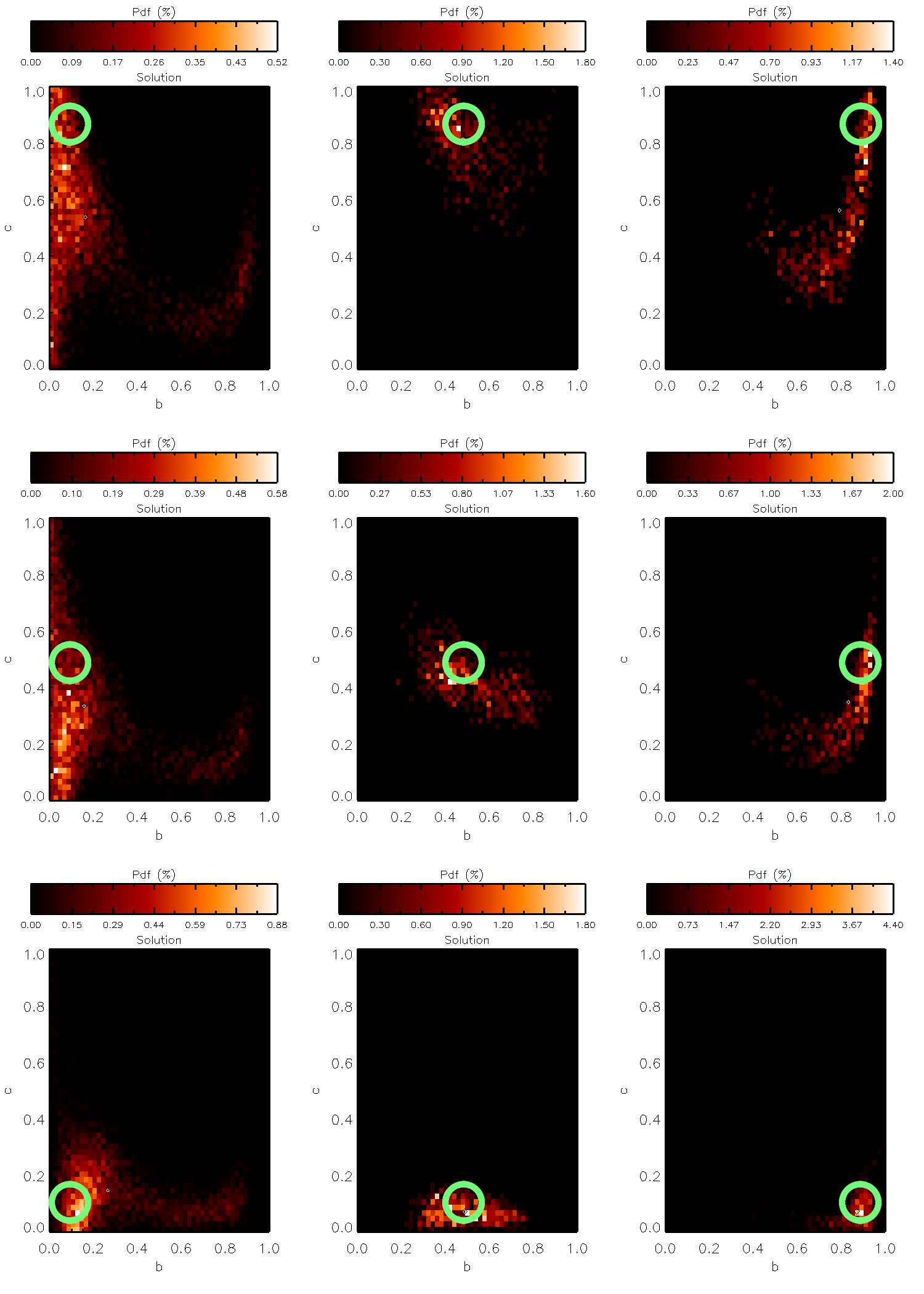}\centering\protect\caption{Same as fig. \ref{fig:Results_EPF_Az0_geom2} but with azimuth=$\left\{ 30^\circ,210^\circ\right\} $.}
\label{fig:Results_EPF_Az30_geom6} 
\end{figure}

\begin{figure}
\includegraphics[bb=0bp 0bp 1500bp 2100bp,clip,height=0.9\textheight]{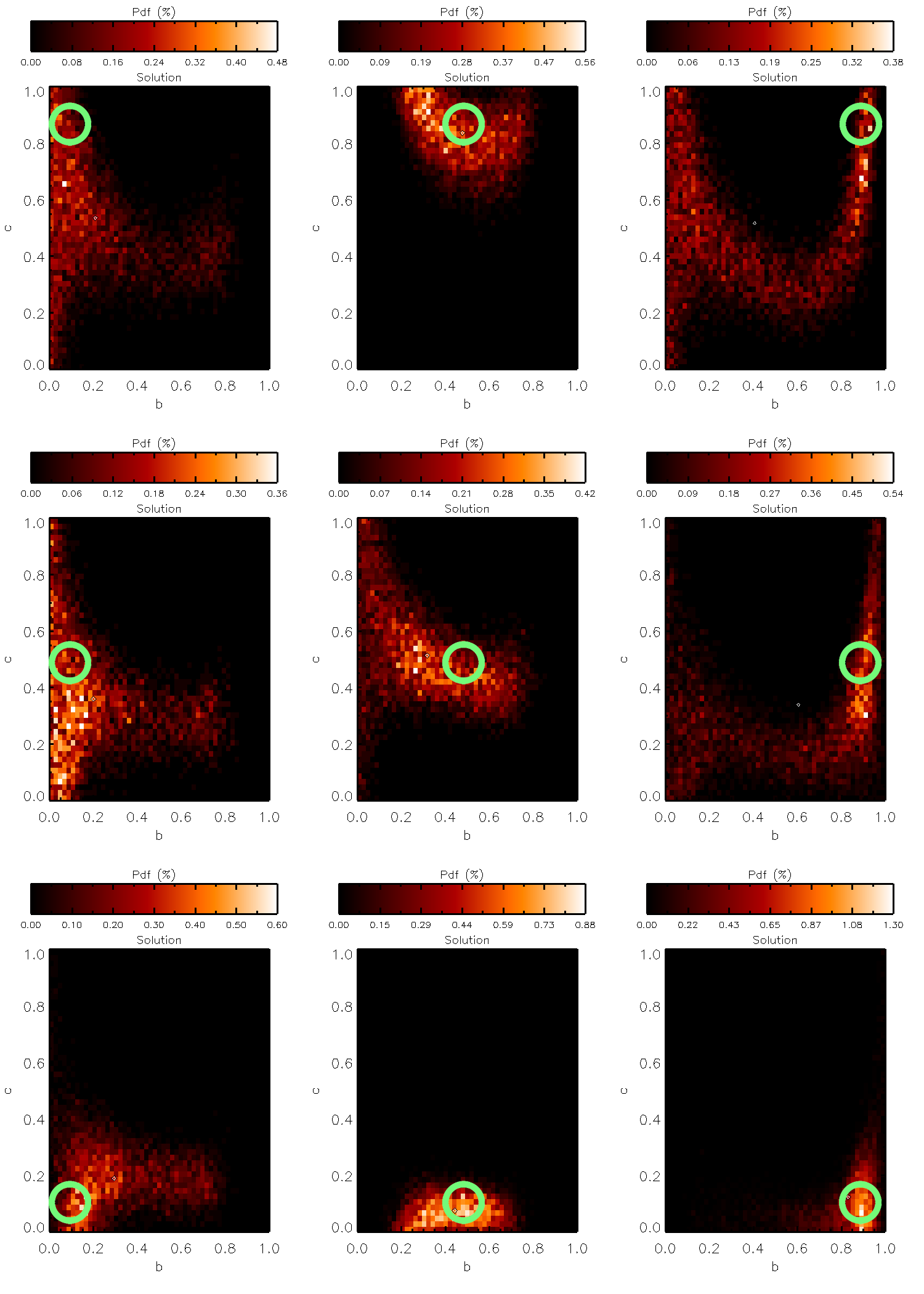}\centering\protect\caption{Same as fig. \ref{fig:Results_EPF_Az0_geom2} but with azimuth=$\left\{ 45^\circ,225^\circ\right\} $.}
\label{fig:Results_EPF_Az45_geom5} 
\end{figure}

\begin{figure}
\includegraphics[bb=0bp 0bp 1500bp 2100bp,clip,height=0.9\textheight]{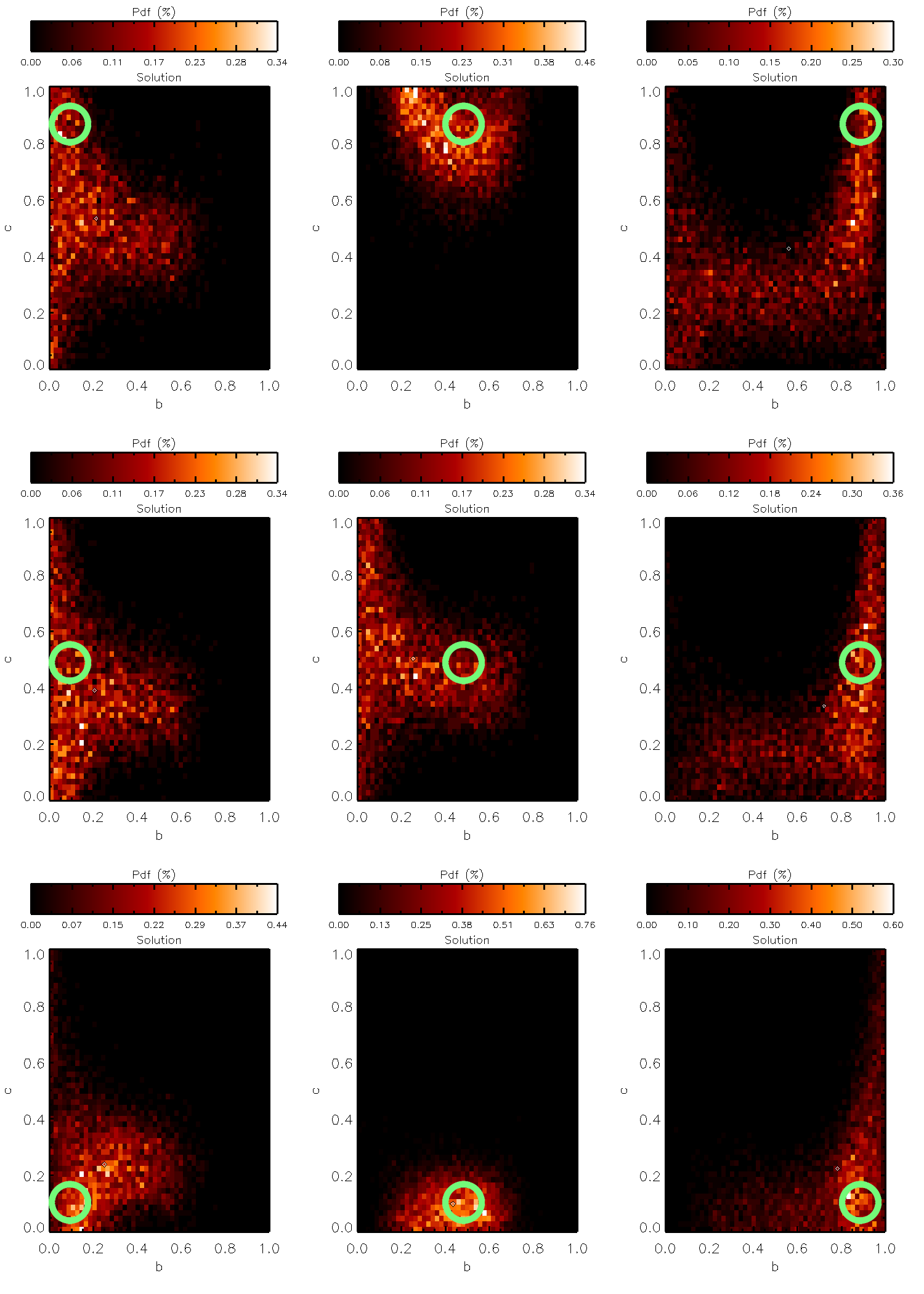}\centering\protect\caption{Same as fig. \ref{fig:Results_EPF_Az0_geom2} but with azimuth=$\left\{ 60^\circ,240^\circ\right\} $.}
\label{fig:Results_EPF_Az60_geom7} 
\end{figure}

\begin{figure}
\includegraphics[bb=0bp 0bp 1500bp 2100bp,clip,height=0.9\textheight]{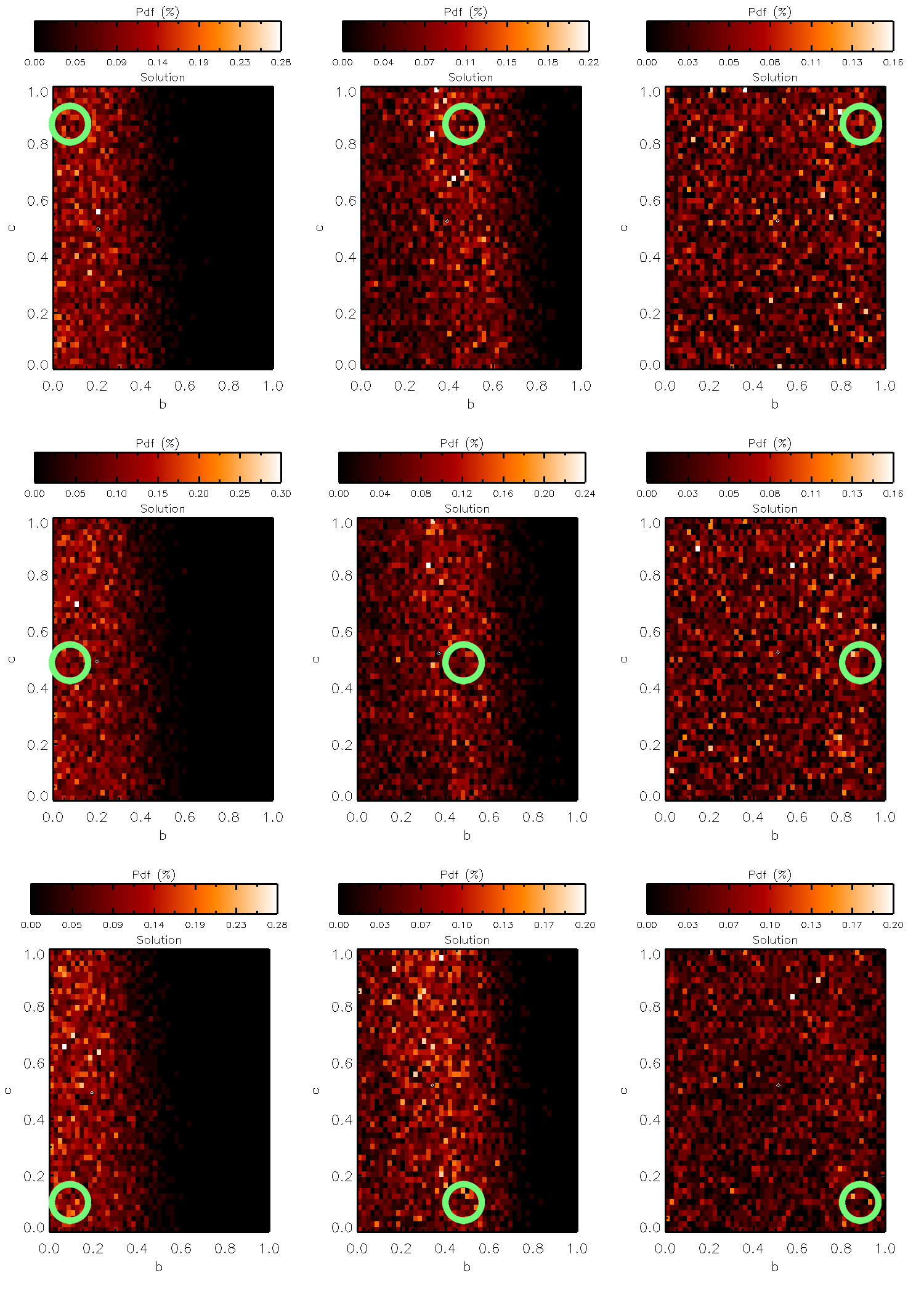}\centering\protect\caption{Same as fig. \ref{fig:Results_EPF_Az0_geom2} but with azimuth=$\left\{ 90^\circ,270^\circ\right\} $.}
\label{fig:Results_EPF_Az90_geom8} 
\end{figure}

Figures \ref{fig:Results_EPF_Az0_geom2}, \ref{fig:Results_EPF_Az30_geom6},
\ref{fig:Results_EPF_Az45_geom5}, \ref{fig:Results_EPF_Az60_geom7}
and \ref{fig:Results_EPF_Az90_geom8} present the results of the favorable
EPF with an azimuthal plane, respectively $\left\{ 0^\circ,180^\circ\right\} $,
$\left\{ 30^\circ,210^\circ\right\} $, $\left\{ 45^\circ,225^\circ\right\} $,
$\left\{ 60^\circ,240^\circ\right\} $, $\left\{ 90^\circ,270^\circ\right\} $.
As expected, the information included in one EPF observation to constrain
the parameters $b$ and $c$ is decreasing by increasing the azimuthal
plane angle. At an azimuthal plane angle of 0$^\circ$ (principal
plane, shown in fig. \ref{fig:Results_EPF_Az0_geom2}), the solutions
are well constrained when the parameter $b$ value is greater or equal
0.5 as observed in section \ref{sub:Principal-plane-EPF}. At an azimuthal
plane angle as low as 30$^\circ$ (Figure \ref{fig:Results_EPF_Az30_geom6}),
the solution is not well constrained in most cases, except for the
cases when $b$=0.5 and the case when $b$=0.9 and $b$=0.1. At an
azimuthal plane angle of 90$^\circ$ (Figure \ref{fig:Results_EPF_Az90_geom8}),
only the parameter $b$ can be qualitatively estimated (i.e., by discriminating
the broad and the narrow scattering), the parameter $c$ is unconstrained
(i.e., all the solutions for the parameter $c$ are possible), because
no phase angles greater than 90$^\circ$ are available in the
data, very important to distinguish the forward and the backward scattering
direction.

Interestingly, the behavior of a high $b$ value coupled with moderate
to high $c$ value differs from the other cases: a ``U'' shape solution
is expressed only in this quadrangle. If the tool used for the inversion
is based on a root mean square minimization, it clearly depends on
the initialization. If the initialization is in the hockey stick shape,
it would converge to a fake local maximum. This effect may be a significant
bias on the estimation of the parameters $b$ and $c$, leading to
an artificial hockey stick effect, present for EPF data with azimuthal
plane larger than 30$^\circ$ and inappropriate inversion method.

\subsection{Effect of noise level on EPF}

\begin{figure}
\includegraphics[bb=0bp 0bp 2000bp 700bp,clip,width=1\textwidth]{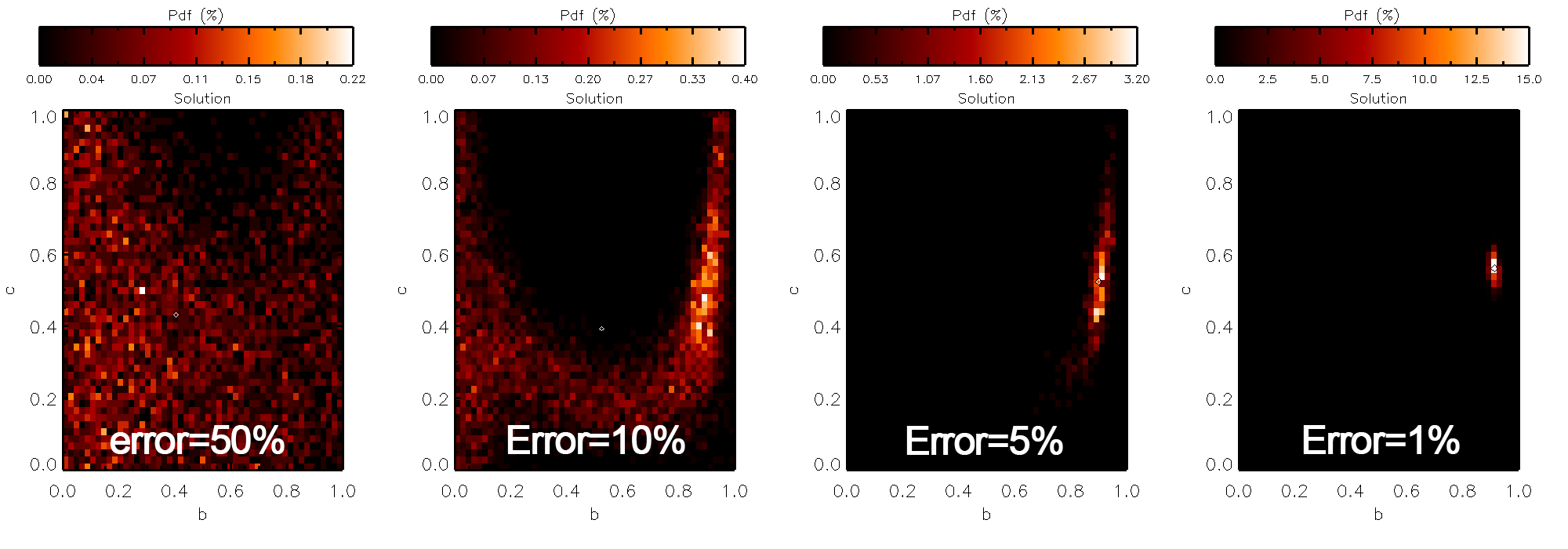}\centering\protect\caption{Probability density function of the parameter $b$ and $c$ using
a favorable EPF observation (17 emergence configurations: 80$^\circ$ ,
70$^\circ$ , 60$^\circ$ , 50$^\circ$ , 40$^\circ$ , 30$^\circ$ ,
20$^\circ$ , 10$^\circ$ , 0$^\circ$ , -10$^\circ$ , -20$^\circ$ ,
-30$^\circ$ , -40$^\circ$ , -50$^\circ$ , -60$^\circ$ ,
-70$^\circ$ , -80$^\circ$ ), incidence=75$^\circ$ , azimuth=$\left\{ 45^\circ,225^\circ\right\} $
and using the following parameter set: $\omega$=0.9, $\bar{\theta}=1^\circ$ ,
$h$=0, $B_{0}$=0, $b$=0.9 and $c$=0.6. Different uncertainties
values are tested: 50\%, 10\%, 5\%, 1\%. The black/white diamonds
represent the average of the PDF.}
\label{fig:Results_EPF_Az45_geom5error} 
\end{figure}

In order to test the noise level on the retrieved parameters, we vary
the noise level (eq. \ref{eq:NoiseLevel}), from 50\% to 1\%. First,
figure \ref{fig:Results_EPF_Az45_geom5error} clearly shows that with
50\% data uncertainty, all solutions in the parameters $b$ and $c$
spaces are possible, thus the parameters are not constrained due to
the large data uncertainty. With 10\% data uncertainty, the solution
is only restricted to the ``U'' shape leading to the artificial
hockey-stick like shape. The $b$ and $c$ couple can be correctly
evaluated with a data uncertainty lower than 5\%. Such uncertainty
level can often be obtained in laboratory measurements \citep{Pommerol_PhotometricpropertiesMars_JGRP2013,Johnson_Spectrogoniometryandmodeling_I2013}
but sometimes also in spaceborne measurements \citep{Fernando_SurfacereflectanceMars_JoGRP2013}.
Finally with 1\% data uncertainty, the solution is well constrained
close to the expected solutions.

The artificial hockey stick effect discussed in the previous section
is only present for uncertainties larger than 5-10\%.

\subsection{Full BRDF\label{sub:HockeyStick_Full-BRDF}}

We test the BRDF observation, sampled 48 times at 2 incidence angles
40$^\circ$ and 60$^\circ$ , 8 emergence angles: 70$^\circ$ ,
50$^\circ$ , 30$^\circ$ , 10$^\circ$ , -10$^\circ$ , -30$^\circ$ ,
-50$^\circ$ , -70$^\circ$ and 3 azimuth angles: 0$^\circ$ ,
45$^\circ$ and 90$^\circ$ . This sampling corresponds to the
combination of standard EPF (table \ref{tab: angular configuration},
case \#2) for 3 azimuthal plane angles, 0$^\circ$ , 45$^\circ$ ,
90$^\circ$ and two incidence angles, allowing high diversity
of geometries. It represents typical laboratory measurements or the
best expected combination of single EPF on the same site. Such combination
of EPF have already been proposed in the literature \citep{Jehl_GusevPhotometry_Icarus2008,Fernando_SurfacereflectanceMars_JoGRP2013,Fernando_Characterizationandmapping_I2015}.

\begin{figure}
\includegraphics[clip,height=0.9\textheight]{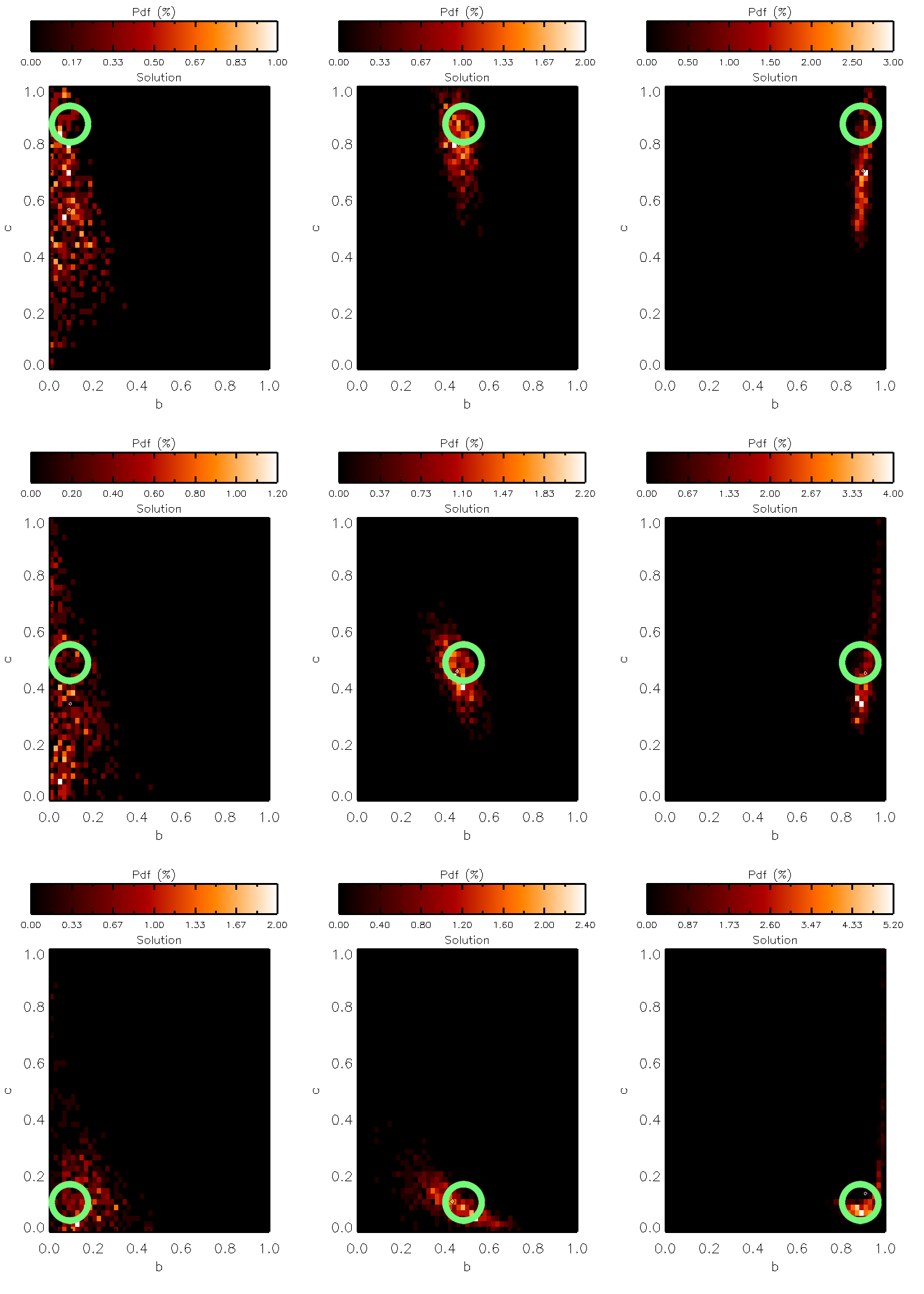}\centering\protect\caption{Same as fig. \ref{fig:Results_EPF-CRISM_Az0_geom1} but for a full
BRDF observation including 48 geometries (2 incidence angles: 40$^\circ$
and 60$^\circ$ , 8 emergence angles: 70$^\circ$ , 50$^\circ$ ,
30$^\circ$ , 10$^\circ$ , -10$^\circ$ , -30$^\circ$ ,
-50$^\circ$ , -70$^\circ$ along 3 azimuth angles: 0$^\circ$ ,
45$^\circ$ , 90$^\circ$ and using the following parameter
set: $\omega$=0.9, $\bar{\theta}$=1$^\circ$ , $h$=0, $B_{0}$=0,
data uncertainties 10\% and 9 combinations of $b$=0.1/0.5/0.9 and
$c$=0.1/0.5/0.9. }
\label{fig:Results_BRDF_geom9} 
\end{figure}

\begin{figure}
\includegraphics[clip,height=0.9\textheight]{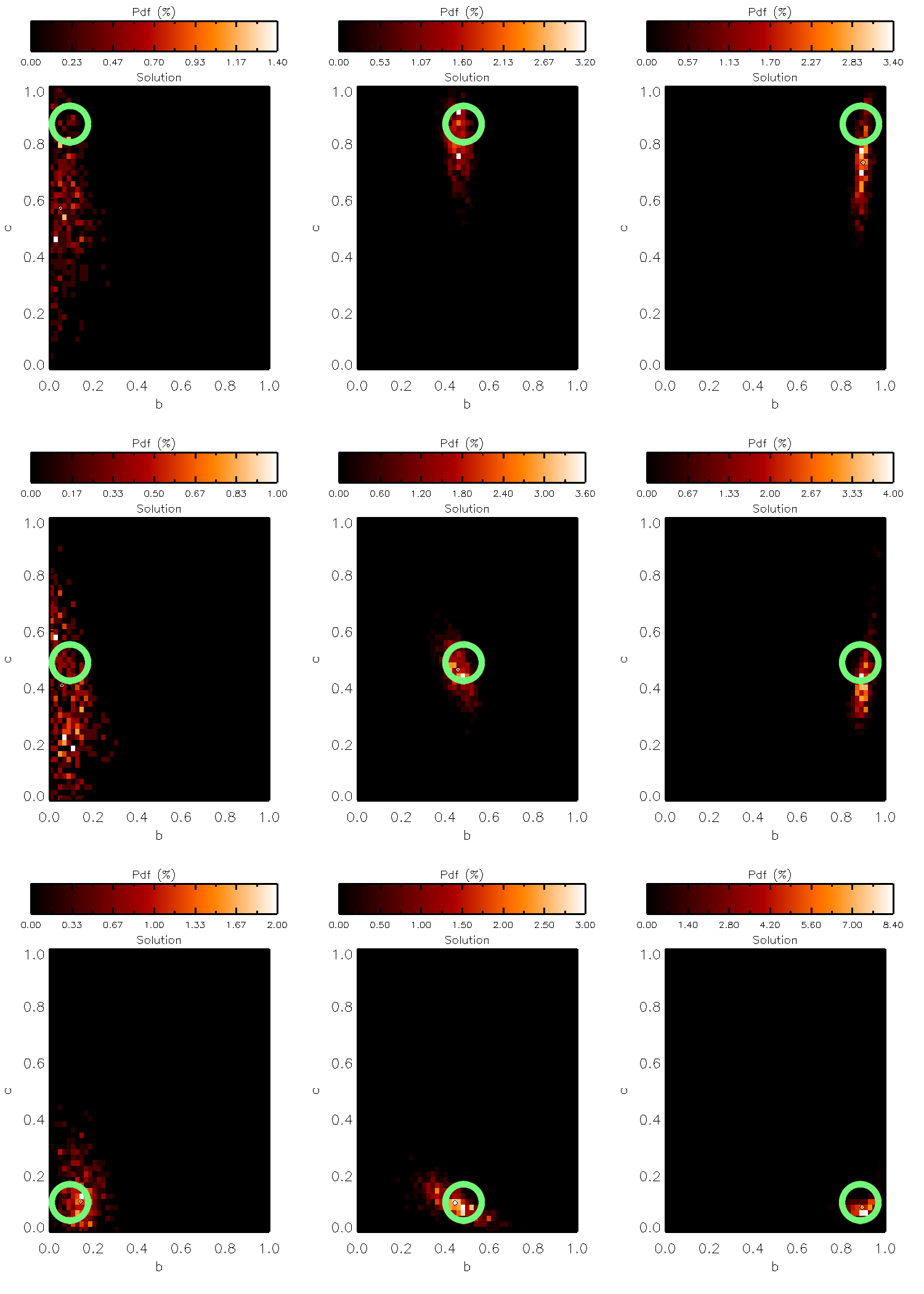}\centering\protect\caption{Same as fig. \ref{fig:Results_BRDF_geom9} but using the Fast Monte
Carlo method explained in section \ref{sub:Fast-MCMC}.}
\label{fig:Results_BRDF_geom9invoptim} 
\end{figure}

The results, presented in fig. \ref{fig:Results_BRDF_geom9}, show
that the BRDF configuration contains enough information to constrain
the parameters $b$ and $c$, in comparison with single EPF (fig.
\ref{fig:Results_EPF-CRISM_Az0_geom1}) with lower uncertainties.
Interestingly, the favorable EPF condition (fig. \ref{fig:Results_EPF_Az0_geom2})
is able to better constrain the phase function parameters than the
standard BRDF presented here. This effect is most probably due to
the maximum phase angle that is limited to 130$^\circ$ for the
BRDF considered here but goes to 155$^\circ$ for the favorable
EPF.

A full BRDF observation at 10\% uncertainties is able to constrain
the parameters $b$ and $c$, and should reduce the contribution to
the hockey stick artifact.

\section{Discussions and Conclusion}

We proposed a rigorous inversion scheme to estimate Hapke model parameter
from measurements, using Bayesian Monte Carlo strategy. The typical
computation time on a 2.5 GHz Intel Core I5, 8Go RAM, is one minute
for a single EPF but can reach one hour for a BRDF. In order to speed
up the convergence, we developed a Fast Monte Carlo strategy, reducing
the computation time by a factor of 100. This strategy is only suitable
on full BRDF or favorable EPF, when gaussian like solutions are expected.

We explored various conditions on synthetic examples: EPF type observations,
one single observation, BRDF type observations in order to study the
propagation of errors from measurements to the parameter space with
10\% uncertainties as a nominal condition. The major conclusions of
this work are:
\begin{itemize}
\item Non-linearities in the Hapke model are important for EPF type measurements
leading to potential multiple solutions, at least with data uncertainties
larger than 5\% and large azimuthal plane angle ($> 30^\circ$).
\item Azimuthal plane in a EPF observation is the most important parameter
to constrain the photometric parameters: the closer to the principal
plane, the best the results. A departure of only $30^\circ$ in
azimuthal plane leads to significant increase of uncertainty.
\item Incidence angle is very important to constrain the parameters in a
EPF, especially the roughness parameter $\bar{\theta}$. A recommendation
for laboratory or spaceborne observation is to sample the highest
incidence angle possible.
\item One single EPF type observation with very favorable conditions (i.e.,
principal plane, incidence at 75$^\circ$ , emergence angle sampling
up to 80$^\circ$ ) is enough to constrain $\omega$, $b$, $c$,
and $\bar{\theta}$ parameters, even with a data uncertainty level
of 10\%.
\item For data uncertainty less than 5\%, the parameters can be estimated
using single EPF under certain geometric configurations: close to
the principal plane (azimuthal angle less than 45$^\circ$ ) and
high incidence angles (greater than 50$^\circ$ ) leading to a broad
phase angle range, containing low and high phase angles, to sufficiently
describe the shape of the photometric curve.
\item In the case of one single observation, with each pixel considered
as a facet with known geometry, $\omega$ can be retrieved with small
uncertainties (1$\sigma$ uncertainties $\leq0.1$ ), but also $b$,
$c$ and in some extent $\bar{\theta}$, in the case of very extreme
phase function. This case is equivalent to one disk resolved image
of a planetary body, assumed to be homogeneous in surface properties.
\item Full BRDF observations allowing a high diversity of geometric sampling
are the best configurations to constrain all the parameter set: $\omega$,
$b$, $c$ and $\bar{\theta}$ . This geometric conditions can be
easily reproduced in laboratory and can be obtained by combining different
EPF observations at varied illumination conditions and/or varied azimuthal
planes. Nevertheless, even BRDF measurements are limited by the phase
range. Our results indicate that a favorable EPF with higher phase
range than a BRDF is better to constraint the parameters.
\item A favorable EPF measurement (i.e., principal plane, incidence at 75$^\circ$ ,
emergence angle sampling up to 80$^\circ$ ) is better to constrain
the parameters than a standard EPF (incidence at 40$^\circ$ and
60$^\circ$ , emergence up to 70$^\circ$ ), most probably due
to better high phase angle sampling.
\item The hockey stick relationship on the $b$ vs $c$ diagram may be the
result of the non-linearity of the Hapke model if the data are from
a EPF type observation and the inversion strategy is based on simple
mean square minimization. However, the Full BRDF type observations
are not biased by the non-linearity. Because the data used in the
Hapke (2012) synthesis are generally BRDF type observations \citep{Hapke_Bidirectionalreflectancespectroscopy_I2012},
it is unlikely that the hockey stick relationship is an artifact from
the inversion method. This confirms that surface material with strongly
backward scattering with narrow lobe may not exist in the nature.
\end{itemize}
Future work should include real laboratory spectra and datasets on
planetary bodies with prioritization using the conclusion of this
study. Also the wavelength dependance of all parameters should be
addressed. Finally, the latest developments of the Hapke model should
be included within this inversion strategy in order to compare the
actual properties of the granular material and retrieved photometric
parameters, given precise uncertainties.

\subsubsection*{Acknowledgement}

We acknowledge support from the ``Institut National des Sciences
de l'Univers" (INSU), the ``Centre National de la Recherche
Scientifique" (CNRS) and ``Centre National
d'Etude Spatiale" (CNES) and through the ``Programme
National de Plan{\'e}tologie"and MEX/OMEGA Program.

\bibliographystyle{elsarticle-harv}

\end{document}